\documentclass[a4paper,11pt]{article}
\usepackage[UKenglish]{babel}
\usepackage[margin=3.0cm]{geometry}
\usepackage{float}
\usepackage[hypertexnames=false]{hyperref}
\usepackage[noend]{algpseudocode}
\usepackage{algorithm}
\usepackage{amsmath,amssymb,mathtools}
\usepackage{amsthm}
\usepackage{cleveref}
\usepackage[toc,page]{appendix}
\usepackage{graphicx}
\usepackage{siunitx}
\usepackage{cite}
\usepackage{pgfplots}
\usepackage{subcaption}
\usepackage{chngpage}
\pgfplotsset{compat=1.7}

\usepackage{authblk}

\def\ln{{\mathrm ln}}
\def\Ln{{\mathrm Ln}}
\def\arg{{\mathrm arg}}
\def\Arg{{\mathrm Arg}}

\def \r {{\bf r}}
\def \rr {{\bf\overline{r}}}
\def \rn {{\bf {r_n}}}

\def \rm {{\bf {r_m}}}
\def \rnz {{\bf {r_N^{\zeta}}}}
\def \dxx {d \overline{x}}
\def \xx {\overline{x}}
\def \zz {\overline{z}}

\def\psiscat{{\psi^s}}
\def\psiscatk{{\psi_j^s}}
\def\psidash{{\psi'}}
\def\psidashk{{\psi'_j}}
\def\psiinc{{\psi^{i}}}
\def\psiinck{{\psi_j^{i}}}
\def\psik{{\psi_j}}

\def\Psiinck{{\bf{\Psi}}_j^{i}}
\def\Psidashk{{\bf{\Psi'}}_j}
\def\Psik{{\bf{\Psi}}_j}
\def\AD{{A^D_j}}
\def\AN{{A^N_j}}

\renewcommand{\bold}[1]{{\bf{#1}}}
\newcommand{\norm}[1]{\left\lVert#1\right\rVert}

\graphicspath{{../Figs/}}
\title{Reconstruction of rough surfaces from a single receiver at grazing angle}
\author[1]{Yuxuan Chen\thanks{\url{chenyx@suda.edu.cn}}}
\author[2]{Mark Spivack \thanks{\url{ms100@cam.ac.uk}}}
\author[2]{Orsola Rath Spivack \thanks{\url{or100@cam.ac.uk}}}

\affil[1]{School of Mathematical Sciences, Soochow University, China}
\affil[2]{Department of Applied Mathematics and Theoretical Physics,
  University of Cambridge, UK} 
\begin{document}
\date{}
\maketitle
%-----------------------------------------------------------------------------

\begin{abstract}
  The paper develops a method for recovering a one-dimensional rough
  surface profile from scattered wave field, using a single receiver and
  repeated measurements when the surface is moving with respect to
  source and receiver.  This extends a previously introduced marching
  method utilizing low grazing angles, and addresses the key issue of
  the requirement for many simultaneous receivers. The
  algorithm recovers the surface height below the receiver point
  step-by-step as the surface is moved, using the parabolic wave integral
  equations.
  Numerical examples of reconstructed surfaces demonstrate that the
  method is robust in both Dirichlet and Neumann boundary conditions,
  and with respect to different roughness characteristics
  and to some degree of measurement noise.
 
\end{abstract}
%-----------------------------------------------------------------------------
\section{Introduction}
\label{sec:introduction}

The calculation of wave scattering by rough surfaces arises in
numerous physical problems 
\cite{voronovich2013wave,Warnick,DESANTO2010139,
  Fung:85,liang2009electromagnetic,Jordan:93}, and
the inverse problem of recovering rough surfaces is a crucial goal in a
wide range of engineering applications, 
ranging from geophysical remote sensing and ocean acoustics (including
under-ice monitoring) 
to nondestructive testing
\cite{vertesy2020analysis,CHOI201827}. 
Many works have focused on rough surface reconstruction via scattered data, e.g.
\cite{retov, Wang:08,
  bao2016shape, zhang1, zhang2, zhang3, newton1, newton2, newton3,
  newton4, CHOI201827, ali1, ali2, ali3, DOLCETTI2021115902,
  10.1121/10.0013506, cnn1, Wombell:91, lines2005time, jyq,
  de1995surface}. 
Such approaches include Kirchhoff \cite{Wombell:91}, Rytov
approximation \cite{retov}, iterative \cite{ali1,ali3} and in
particular Newton methods 
\cite{newton1,newton2,newton3,newton4}, linear sampling \cite{zhang2,zhang3},
and more recently deep learning \cite{cnn1}. Most of these use single
frequency data while others have utilized multiple frequencies
\cite{bao2016shape}.   
In particular, in \cite{rough1,rough2,rough3,spivack1,spivack2},
efficient reconstruction algorithms have been developed under the
regime of grazing angle incidence 
using the parabolic wave equation.
When the incident angle is small, and forward propagation
dominates, the Helmholtz equation can be approximated by the
parabolic wave equation.
This allows the boundary integral equations to be replaced by the
Parabolic Integral Equations (PIE)
\cite{10.1121/1.2024537}, using which the computational cost of both
direct and inverse problems can be significantly reduced.
However a drawback of some of the above methods is that in practical
applications the scope for number of receivers may be limited, and
indeed their placement may interfere with the scattered field to be measured.

One approach to this issue is to allow the measurement
apparatus to move with respect to the surface (or equivalently fixed
source and receiver with a moving surface) and take repeated
measurements using fewer receivers.   Such a situation is feasible for
applications such as NDT of surface roughness, or under-ice monitoring.
However, this poses additional complexity for the computational implementation.

In this paper, an algorithm is developed to reconstruct the rough
surface which is moving, for scattering at the grazing angle. The
algorithm relies on the scattered data at a single fixed point, thus
only one receiver is needed for the reconstruction.  This approach is
analogous to previous works \cite{rough1,rough2} based on PIE,  but in
those cases, $N$ receivers were required where $N$ is number of
unknowns determined  by the discretisation size.  Both the Dirichlet
(TE incident field) 
and Neumann (TM) boundary conditions are treated here.
In this marching approach, the surface profile
is recovered point by point as the surface moves. 
We first recover the surface current
with which the surface height below
the receiver point can then be reconstructed using the coupled integral
equations.

A variety of numerical examples are presented to validate the proposed
method.  The source is taken by the Gaussian beam with a single
frequency.  Two types of rough surfaces are tested: Gaussian and
sub--fractal.
It is found that the proposed method can successfully
capture the detailed features of the rough surface.  The method is 
tested with noisy data; although this introduces fine-scale
oscillations, reconstruction still agrees well with the actual
surface, and these oscillations can easily be filtered out.  The
error between the reconstructed and actual surfaces in the
$\ell_2$--norm is measured with respect to the discretisation size,
surface type, surface scale, and the receiver height.

The layout of the paper is structured as follows. In
\cref{sec:background}, we give a brief overview of the direct problem
using parabolic
integral equations. The problem setting and reconstruction algorithm
are described in \cref{sec:reconstruction}, which includes the
detailed formulations for both the Dirichlet and Neumann boundary
conditions.  Numerical tests are carried out in \cref{sec:tests},
where the method is examined by both the error and the performance
with noisy data. Finally, concluding remarks and future work are
discussed in \cref{sec:conclusions}.

%-----------------------------------------------------------------------------
\section{Mathematical Background}
\label{sec:background}

Boundary integral formulations of rough surface scattering problem
have been extensively studied, for the full Green's function
(Helmholtz regime), for example in \cite{Warnick,DESANTO2010139}, and
for the parabolic integral equations (PIE) on which the
current study is based \cite{Tappert1977,10.1121/1.2024537,10.1121/1.399327}.
We briefly review the assumptions and the governing integral equations for
the parabolic regime studied here.
Consider a scattering problem in which an
electromagnetic field is incident at a low grazing angle on a perfectly
conducting corrugated surface varying in one dimension only, so that
the problem becomes two-dimensional and scalar.
The incident field is taken to be time-harmonic and either
horizontally or vertically plane polarized, i.e. transverse electric (TE) or
transverse magnetic (TM).  Time
dependence may be suppressed so that we consider the time-reduced
component.
The coordinate axes are $x$ and $z$, where $x$ is horizontal,
and $z$ is vertical. We denote the time--harmonic wave field as $E(x,z)$.
The governing equation for the wave field is the Helmholtz equation,
\begin{equation}
\nabla^2 E(x,z) + k^2 E(x,z) = 0,
\end{equation}
where $k$ is the wavenumber.
If the incident angle is small enough so that the forward
propagation dominates, then there is a slowly varying component of the
wave field, denoted by $\psi(x,z)$, given by
\begin{equation}
\psi(x,z) = E(x,z) \exp(-ikx).
\end{equation}
With this approximation, the backscattered component
is neglected, and the slowly varying wave field
can be governed by the parabolic wave equation:
\begin{equation}
\frac{\partial \psi}{\partial z} =
\frac{i}{2k}\frac{\partial^2 \psi}{\partial x^2}.
\end{equation}
The Green's function for the parabolic wave equation is
in the form of
\begin{equation} G(x,z; \xx, \zz) = \frac{1}{2}\sqrt{\frac{i}{2\pi k
(x-\xx)}} \exp{\left[\frac{ik(z-\zz)^2}{2(x-\xx)}\right]},
\end{equation}
when $\xx<x$, and $G = 0$ otherwise \cite{10.1121/1.2024537,
10.1121/1.399327}.
We denote the incident wave field, the scattered wave field, and
the total wave field as $\psiinc(x,z)$, $\psiscat(x,z)$, and $\psi(x,z)$,
respectively, so that $\psi = \psiinc + \psiscat$.

The forward problem (to obtain the scattered field from a known
surface) is typically solved via the Kirchhoff--Helmholtz equations
which relate the scattered and incident fields via the surface
currents ($J$ or $K$ depending on polarization).
The inverse problem tackled below is the reconstruction of a rough
surface from scattered data using a single receiver, by allowing the
surface to move with respect to the source and receiver.
Consider first an arbitrary rough surface $z=h(x)$
on the domain $[0,L]$.
The source will be a Gaussian beam propagating at a relatively small
angle to the horizontal, so we can consider the slowly varying
component, and replace the Kirchhoff--Helmholtz equation by the
coupled parabolic integral equations relating the incident field and
the scattered field~\cite{10.1121/1.399327}.
Let the space $V$ contain all the points lying above the surface with
$V = \{(x,z): z > h(x)\}$,
and let the space $S$ contain the points lying on the surface with
$S = \{(x,z): z = h(x)\}$.  Two boundary conditions are considered in
this paper, Dirichlet and Neumann. If we assume the Dirichlet boundary
condition, which corresponds to a transverse electric (TE) field
impinging on a perfectly conducting surface, then the coupled integral
equations are
\begin{equation}
\psiinc(\r) = -\int_{0}^{x} \frac{\partial
\psi(\rr)}{\partial \zz} G(\r; \rr) \, \dxx, \quad \quad \r, \rr \in
S,
\label{eqn:d1}
\end{equation}
where points $\r$ and $\rr$ both lie on the surface,
and
\begin{equation}
\psiscat(\r) = \int_{0}^{x} \frac{\partial
\psi(\rr)}{\partial \zz} G(\r; \rr) \, \dxx, \quad \quad \r \in V, \rr
\in S,
\label{eqn:d2}
\end{equation}
where $\rr$ again lies on the surface, and
 $\r$ is any point above the surface.  For simplicity, we
denote $\psidash := \partial \psi / \partial z$.  On the other hand, if
we assume the Neumann boundary condition, which corresponds to a
transverse magnetic (TM) polarized field or an acoustically hard surface,
then the coupled integral equations become
\begin{equation} \psiinc(\r) = \frac{\psi(\r)}{2} + \int_{0}^{x}
\psi(\rr) \frac{\partial G(\r; \rr)}{\partial \zz} \, \dxx, \quad
\quad \r, \rr \in S,
\label{eqn:n1}
\end{equation}
where points $\r$ and $\rr$ both lie on the surface,
and
\begin{equation} \psiscat(\r) = -\int_{0}^{x} \psi(\rr) \frac{\partial
G(\r; \rr)}{\partial \zz} \, \dxx, \quad \quad \r \in V, \rr\in S,
\label{eqn:n2}
\end{equation}
where the point $\rr$ still lies on the surface, but
point $\r$ can be any point in the medium.

%-----------------------------------------------------------------------------
\section{Reconstruction of moving surface}
\label{sec:reconstruction}
%------------------------------------------------------------------------------
\subsection{Problem setting}
\begin{figure}
\centering
\includegraphics[width=0.5\linewidth]{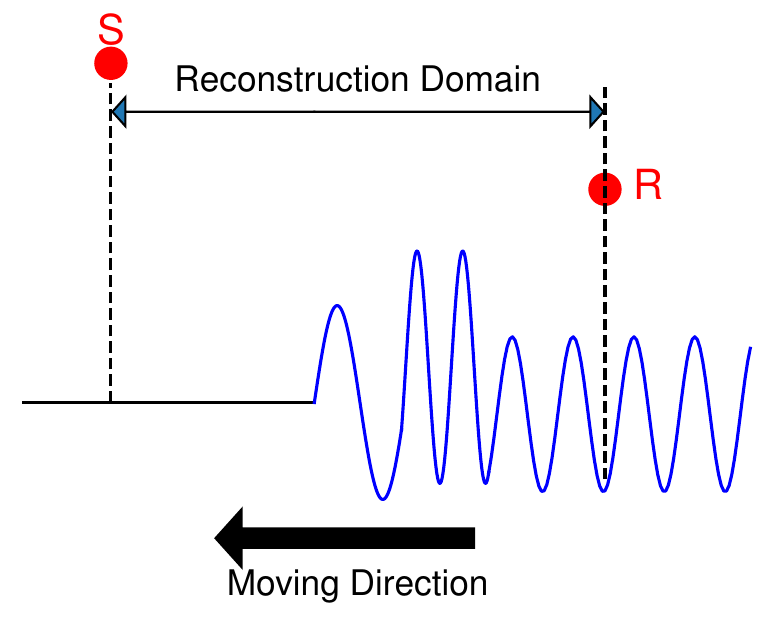}
  \caption{Problem setting for reconstruction of rough surface
  moving with respect to source and receiver,
  in which the source (S) and the receiver (R) are
  located at $x=0$ and $x=L$, respectively.
  The reconstruction domain is the region between the source and
  receiver planes.}
  \label{fig:problem_setting}
\end{figure}

This paper deals with the situation where
the surface moves horizontally in the negative $x$ direction with
respect to a fixed source and receiver  (see
\cref{fig:problem_setting}).
This is obviously equivalent to source and receiver moving over a
static surface, and the formulation here is chosen for convenience.

The rough segment to be recovered is considered as part of an extended
surface with a flat part and rough
part. For simplicity the length of each of these regions
is set to be $L$, making the total length of the surface $2L$.
The source and receiver are above the surface, at given heights and separated 
horizontally by distance $L$. Define the {\it{reconstruction domain}} to be
the region 
lying between the vertical
planes of source and receiver, $x=0,L$ respectively. An illustrative figure
of the physical setting is shown in Fig. \ref{fig:problem_setting}.

Thus the surface profile $h(x)$ is initially flat for $x \le L$ and
irregular for $x \ge L$. The region $[0,2L]$ is discretized
using $2N+1$ equally spaced points
$x_n = n\Delta x$ for $n=0,1,\cdots,2N$ where $\Delta x = L/N$.
The surface values at initial time in the area $[0,2L]$ can be thought
of the discretized values $H_q := h(x_{N+q})$ for
$q=-N, \cdots, 0,1,\cdots,N$, in which
negative $q$ corresponds to the flat part, and positive $q$ corresponds
to the rough part of the extended surface.
There are in total $N + 1$ time steps $t_0$ to $t_N$.
In the inverse problem from time $t_j$ to $t_{j+1}$, the surface moves by
one spatial distance $\Delta x$ towards the left and
a new measurement of scattered field is taken.

\begin{figure}
  \centering
  \includegraphics[width=0.7\linewidth]{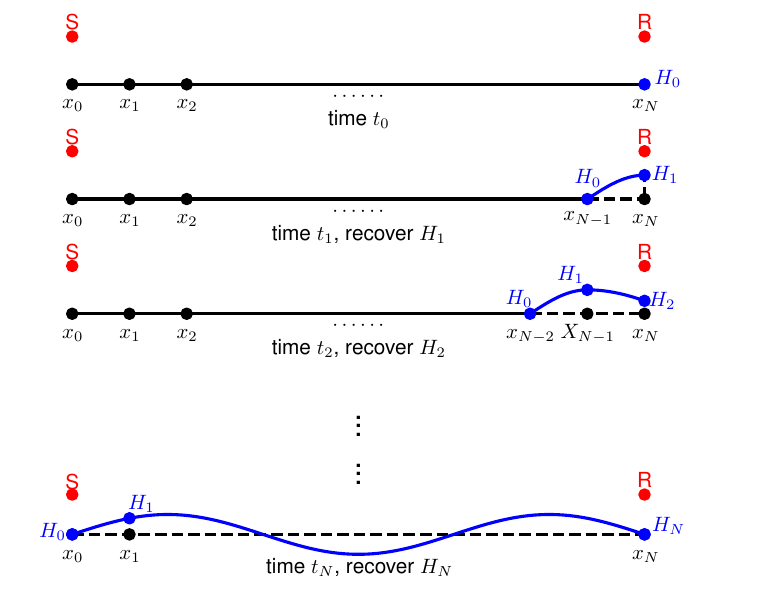}
  \caption{Illustration of reconstruction domain at successive time steps
  lying between source (S) and receiver (R). At time step $t_j$, the surface
  in the reconstruction domain is $h_j(x)$.}
  \label{fig:time_step}
\end{figure}

The algorithm is applied within the reconstruction domain
$[0,L] \equiv [x_0, x_N]$ between the source and receiver.
As surface moves, the surface height depends on the time step.
we denote the surface at the initial time
step $t_0$ as $z=h_0(x)$ with $x\in[0.2L]$.
The repositioned surfaces in the reconstruction domain $[0,L]$
at the time step $t_j$ for $j\ge 1$ is denoted by $h_j(x)$ given by
\begin{equation}
h_j(x) := h_0(x - j \Delta x), \quad \quad x\in[0,L].
\end{equation}
Accordingly, the spaces $V(t_j)$ and $S(t_j)$ represent the points lying above and
on the surface, respectively, at the step $t_j$, which are given by
$V(t_j) = \{(x,z): z > h_j(x), x\in[0,L]\}$,
and
$S(t_j) = \{(x,z): z = h_j(x), x\in[0,L]\}$.
A schematic illustration of successive time steps is shown in
\cref{fig:time_step}.
It is clear that at step $t_{j}$, the last point of
surface in the reconstruction domain is $h_j(L) = h_j(x_N) \equiv H_{j}$.
Finally, we denote by $\psidashk$, $\psi_j$, and $\psiscatk$ the surface
currents and scattered field at time step $t_j$.

%-----------------------------------------------------------------------------
\subsection{Reconstruction in the Dirichlet case}

Assume first that the Dirichlet boundary condition holds.
For the rough surface at the reconstruction domain at step $t_j$,
\cref{eqn:d1} and \cref{eqn:d2} become
\begin{equation}
\psiinc(\r) = -\int_{0}^{x} \psidashk(\rr)
  G(\r; \rr) \, \dxx, \quad \quad \r, \rr \in S(t_j),
\label{eqn:dt1}
\end{equation}
and
\begin{equation}
\psiscatk(\r) = \int_{0}^{x} \psidashk(\rr)
  G(\r; \rr) \, \dxx, \quad \quad \r \in V(t_j), \rr \in S(t_j).
\label{eqn:dt2}
\end{equation}

Returning to the discretized problem, 
surface values $H_n$ for $n\in\{-N,..,0\}$
correspond to the flat segment
which lies initially within the reconstruction interval $[0,L]$.
The algorithm seeks to recover the unknown surface
values $H_n$ for $n\in \{1,...N\}$.
At step $t_1$ all surface values within the interval are known except
that at $x=L$ which is now $H_1 \equiv h_1(x_{N})$. This value is
recovered as described below, and the procedure is 
carried out repeatedly. More generally, at each step
$t_j$,  only the right--most value $H_j$
of those on the interval $[0,L]$ is unknown and is to be recovered.
At each step, the reconstruction algorithm is an extension of the
marching procedure in the previous works~\cite{rough1,rough3} as we now describe.
The method requires a sequence of scattered field measurements
$\psiscatk(\rnz)$ at the fixed point $\rnz = (x_N, \zeta)$ where
$\zeta$ is the height of the receiver for time steps $t_j$, $j=1,2,\cdots,N$.
The algorithm seeks to reconstruct the unknown values to the right
successively as the surface moves.   

At time step $t_j$, the surface values $H_0, \cdots,H_{j-1}$ are known.
For any point $\rn = (x_n, h_j(x_n)) \in S(t_j)$ with $x_n \in [x_1, x_N]$,
\cref{eqn:dt1} is written as a sum of $n$ subintegrals with
\begin{equation}
\psiinc(\rn) = -\sum_{m = 1}^{n} \int_{x_{m-1}}^{x_m}
G(\rn; \rr) \psidashk(\rr) \, \dxx, \quad \quad \rn, \rr\in S(t_j).
\end{equation}
We assume that the surface derivative $\psidash$,
and the Green's function (except in the last summand where it is singular)
vary slowly over the subintervals, and can be treated as constant.
For $n > 1$ the integral formula then becomes
\begin{equation}
\psiinc(\rn) \cong -\sum_{m = 1}^{n-1} G(\rn;
\rm) \psidashk(\rm)  \Delta x -\psidashk(\rn) \int_{x_{n-1}}^{x_n}
G(\rn; \rr) \; \dxx, \quad \quad \rn, \rr\in S(t_j).
\end{equation}
At this point, we assume that the surface value at the last point
is zero, $H_j \equiv h_j(x_n) = 0$.
By assumption all surface values required to evaluate the sum are
known.
If we take $n = 1, 2, \cdots, N$, a linear system can be
generated with
\begin{equation}
\AD {\Psidashk} = {\Psiinck},
\end{equation}
where $\AD$ is a matrix of size $N \times N$ given by
\begin{equation}
\AD(n, m) =
\begin{cases}
-\Delta x G(\rn; \rm), \quad &m<n\\
-\int_{x_{n-1}}^{x_n} G(\rn; \rr) \, \dxx, \quad &m= n
\end{cases}
, \quad \rn, \rm, \rr \in S(t_j),
\end{equation}
for $1 \leq m \leq n \leq N$, and $\Psidashk$ and $\Psiinck$ are
vectors of size $N$ with
\begin{equation}
\begin{aligned}
\Psidashk &= [\psidash_j(\bold{r_1}), \psidash_j(\bold{r_2}),
\cdots \psidash_j(\bold{r_N})]^T,\\
\Psiinck &= [\psiinck(\bold{r_1}),
\psiinck(\bold{r_2}), \cdots \psiinck(\bold{r_N})]^T.
\end{aligned}
\label{eqn:vec}
\end{equation}
By solving the linear system, the surface derivative on
all the discrete surface points over the reconstruction domain can be obtained.
The singularity in the diagonal term of $\AD$ has been treated in \cite[Appendix B]{rough1},
which is given by $\AD(n,n)=2\frac{\alpha}{\gamma}\left[C\left(\gamma \sqrt{\Delta x}\right) + i S\left(\gamma \sqrt{\Delta x}\right) \right]$
where $\gamma = \sqrt{k/2}\lvert h_j'(x_n)\rvert$, and $C$ and $S$ are the cosine and sine Fresnel integrals.
Note that the system $\AD$ is lower--triangular, thus the inversion of this system is
computationally cheap.

We use the second integral equation (\cref{eqn:d2}) to recover the
surface height at the end $h_j(x_N)$.  The receiver is at the point
$\rnz = (x_N, \zeta)$
where $\zeta$ is a fixed value above the rough surface.
Under the assumption that the surface derivative and Green's function can be
viewed as constant over the subinterval,  the scattered field at
the point of receiver at time step $t_j$ can be written by
\begin{equation}
\psiscatk(\rnz)
  = \sum_{m = 1}^{N-1} \Delta x
\psidashk(\bold{r_m}) G(\rnz; \rm) +\psidashk(\bold{r_N})
\int_{x_{N-1}}^{x_N} G(\rnz; \rr)\, \dxx, \quad
\bold{r_N}, \rm, \rr \in S(t_j).
\label{eqn:diri_scatter}
\end{equation}
For the last integral, we approximate the integral
using the midpoint:
\begin{equation}
\int_{x_{N-1}}^{x_N} G(\rnz; \rr) \dxx =
\frac{\alpha}{\sqrt{x_N-\xi_N}} \exp\left[\frac{ik(\zeta -
h_j(x_N))^2}{2(x_N-\xi_N)}\right] \Delta x,
\label{eqn:diri_last_term}
\end{equation}
where $\xi_N = 1/2(x_{N-1} + x_{N})$, and $\alpha =
1/2(i/2\pi k)^{1/2}$. As we know the surface derivative $\psidash$ at all
the discrete points, and all the surface values
other than the last one $h(x_N)$, then the first $(N-1)$ terms in
\cref{eqn:diri_scatter} can be calculated. To recover the surface
height $h_j(x_N)$, the problem becomes solving an exponential
equation. The logarithm of a complex--valued number is ambiguous, for
$z\in \mathbb{C}$, $\log z = \ln(|z|) + i\arg(z)$, where $\arg(z)=
\Arg(z) + 2k\pi$ with $\Arg(z)\in[-\pi,\pi]$ and $k\in
\mathbb{Z}$. However, if we assume that the surface height is small
enough, we can simply take the principal component, namely, we use
$\Ln(z) = \ln |z| + i \Arg(z)$. Combining the two equations
\cref{eqn:diri_scatter} and \cref{eqn:diri_last_term}, the surface
height at the last point is recovered with
\begin{equation}
h_j(x_N) =\zeta - \sqrt{\frac{\Delta x}{ik} \Ln(Q_N)},
\label{eqn:diri_h1}
\end{equation}
where
\begin{equation}
Q_N = \frac{\psiscatk(\rnz)
 - \sum\limits_{m=1}^{N-1}\Delta x \psidashk(\rm) G(\rnz; \rm)}
 {\alpha \sqrt{2 \Delta x} \psidashk({\bf r_N})}, \quad \rm, {\bf r_N}\in S(t_j).
 \label{eqn:diri_h2}
\end{equation}
Finally, at time step $t_{j}$, the last point
surface height is equivalent to $H_{j}$, and we set $H_j :=
h_j(x_N)$.

%-----------------------------------------------------------------------------
\subsection{Reconstruction in the Neumann case}
  
Similar treatment can be applied to the Neumann case.
We still first assume that $H_n = h_0(x_N) := 0$.
Under the assumption that the total surface wave and the derivative
of the Green's function vary slowly over each subinterval and can be treated
as constants, then
at any time step $t_{j}$, \cref{eqn:n1} can be written by
$N$ subintegrals
\begin{equation}
\psiinck(\rn) = \frac{1}{2}\psik(\rn) - \sum\limits_{m=1}^{n} \psik(\rm)
\int_{x_{m-1}}^{x_m} \frac{\partial G(\rn ; \rr)}{\partial \zz} \,
\dxx, \quad \quad \rn, \rm,\rr\in S(t_j),
\end{equation}
for the points $\rn = (x_n, h_j(x_n))$, $\rm =
(x_m,h_j(x_m))$, and $\rr = (\xx, h_j(\xx))$.  A linear system is then
generated with
\begin{equation}
\AN \Psik= {\Psiinck},
\end{equation}
where $\AN$ is a matrix of size $N \times N$ given by
\begin{equation}
\AN(n,m) =
\begin{cases}
\Delta x \frac{\partial G(\rn ; \rm)}{\partial \zz} \quad \quad &m < n\\
\frac{1}{2} - \int_{x_{n-1}}^{x_n}\frac{\partial G(\rn ;
\rr)}{\partial \zz} \, \dxx \quad \quad &m = n
\end{cases} , \quad \quad \rn, \rm, \rr \in S(t_j),
\end{equation}
for $1 \leq m \leq n \leq N$, and $\Psik$ is the
vector of total wave along surface of size $N$ given by
\begin{equation}
\Psik= [\psik(\bold{r_1}), \psik(\bold{r_2}),\cdots, \psik(\bold{r_N})]^T,
\end{equation}
and ${\Psiinck}$ is the vector of incident wave
along surface given in \cref{eqn:vec}.  The singular integral in the
diagonal term of $\AN$ has been calculated in \cite{rough3}, with
$\AN(n,n)=2\frac{\beta}{\gamma}\left[C\left(\gamma \sqrt{\Delta x}\right) + i S\left(\gamma \sqrt{\Delta x}\right) \right]$,
where $\beta = -i/2\sqrt{ik/2\pi}$.
The surface current, $\Psik$, can be obtained by inverting the low--triangular matrix $\AN$.

We use \cref{eqn:n2} to reconstruct the surface profile to the right, which is $h(x_N)$.
At the point of receiver $\rnz = (x_N, \zeta)$, \cref{eqn:n2} can be
expressed in the form:
\begin{equation}
\psiscatk(\rnz) = \Delta x \sum\limits_{m = 1}^{N-1} \psik(\rm) \frac{\partial G(\rnz ; \rm)}{\partial \zz}
+ \psik(\bold{r_N})\int_{x_{N-1}}^{x_N} \frac{\partial G(\rnz ;\rr)}{\partial \zz} \, \dxx,
\label{eqn:neu_scatter}
\end{equation}
where the terms $\rm$, $\bold{r_N}$, $\rr$ are again evaluated on the surface $S(t_j)$.
As in the Dirichlet case, the first $(N-1)$-sum is
known. We directly approximate the last integral with
\begin{equation}
\int_{x_{N-1}}^{x_N} \frac{\partial G(\rnz ;\rr)}{\partial \zz} \, \dxx
\approx \beta \Delta x \frac{\zeta - h_j(x_N)}{(x_N - \xi_N)^{3/2}}
\exp\left[\frac{ik}{2} \frac{(\zeta - h_j(x_N))^2}{x_N - \xi_N} \right],
\end{equation}
where $\xi_N$ is the midpoint defined by $\xi_N = (x_N
+ x_{N-1})/2$.  Rearrange \cref{eqn:neu_scatter}, and take the
absolute value both sides, we have
\begin{equation}
\left \lvert \frac{\zeta - h_j(x_N)}{(x_N -\xi_N)^{3/2}} \right \rvert =
\left \lvert \frac{\psiscatk(\rnz) - \sum\limits_{m = 1}^{N-1} \Delta x \psik(\rm)
\frac{\partial G(\rnz ; \rm)}{\partial \zz}} {\beta \psik(\bold{r_N}) \Delta x} \right\vert.
\end{equation}
Provided that the receiver is higher than the surface,
the surface height $h_j(X_N)$ can be obtained via
\begin{equation}
h_j(X_N) = \zeta - \left \lvert \frac{W_N \sqrt{\Delta x}}{2 \sqrt{2} \alpha \psik(\bold{r_N})} \right \rvert,
\end{equation} 
where
\begin{equation}
W_N = \psiscatk(\rnz) - \sum\limits_{m = 1}^{N-1} \Delta x \psi(\rm) \frac{\partial G(\rnz ; \rm)}{\partial \zz}.
\end{equation}
Finally, at the time step $t_j$, we set $H_j := h_j(x_N)$.

%-----------------------------------------------------------------------------
\section{Numerical results}
\label{sec:tests}

The performance of the proposed method is examined in this
section.
For ease of comparison with previous PE results, in the
numerical examples the wave number is set to be $k =
1$, resulting in the wavelength of $\lambda = 2 \pi$.
The extended surface is made up of a flat surface and a rough surface,
which in total have length $2L$, where $L = 200$, approximately $31.8\lambda$.
The source is a Gaussian beam, given by
\begin{equation}
\psiinc(x,z) = \frac{w}{(w^2 +
2ix/k)^{1/2}}\exp\left[-\frac{(z-z_0)^2}{w^2+2ix/k} \right],
\end{equation} 
where it is centred at $(x_0,z_0)$ with $z_0 = 22.4$,
and the beam width $w$ is given by $w = 8$.  The scattered data is measured
at the point $(x_N, \zeta)$,
where $\zeta$ is a fixed height.  The grid points
for generating scattered data and surface reconstruction are set
different.  The reconstruction domain is discretized to $N$
subintervals at all time steps.  Noisy data is also used to test
the robustness of the method: If the scattered field measured by the
receiver is $\psiscat$, then the noisy data is given by
\begin{equation}
\psiscat_{\text{noise}} = \psiscat (1 + \epsilon
\vartheta),
\end{equation}
where $\epsilon$ is the noise level and $\vartheta$ is
the random number in $[-1,1]$ generated with Gaussian distribution.

The rough surfaces used in the tests are randomly generated by a given
autocorrelation function (a.c.f.) $\rho(\eta)$, where $\eta = x -
x'$. Two types of rough surfaces are employed here, having Gaussian--type
a.c.f.
\begin{equation} \rho(\eta) = \sigma^2 \exp(\frac{-\eta^2}{l^2}),
\end{equation} 
where $\sigma$ is the variance and $l$ is the surface
scale (autocorrelation length); and `sub--fractal'
a.c.f.
\begin{equation} \rho(\eta) = \sigma^2 \left(1 + \frac{\lvert \eta
\rvert}{l} \right)\exp \left( - \frac{\lvert \eta \rvert}{l} \right).
\end{equation} 
For both cases, we set $\sigma = 0.2$. The
typical peak-to-trough of the tested rough surfaces is around $0.6$,
around $\lambda/10$.
Qualitatively, at fine scales, the Gaussian surface is relatively smooth,
whereas the sub-fractal surface is more jagged.

Within the reconstruction algorithm, all functions are initially
allowed to be complex--valued.
As expected it is found that the reconstructed surface has only a
small imaginary component, which can be neglected.
The discretization size $N$ is varied to test the performance of the method. 
The surface moves in the negative $x$-direction by $\Delta x$ each step
for $\Delta x = L / N$.  The $\ell_2$--norm is used to quantify and examine
the error, with
\begin{equation}
\norm{e}_2 =
\frac{\norm{H-\hat{H}}_{\ell_2}}{\norm{\hat{H}}_{\ell_2}}.
\end{equation} 
where $H$ is the reconstructed surface and $\hat{H}$ is
the exact surface. Note that this can underestimate the
qualitative performance, for example it may exaggerate
the error due to a small horizontal shift in the reconstructed surface.

%-----------------------------------------------------------------------------
\subsection{Dirichlet boundary condition}
\begin{figure}
  \centering
  \begin{subfigure}{0.46\textwidth}
  \centering
  \includegraphics[width=0.95\linewidth]{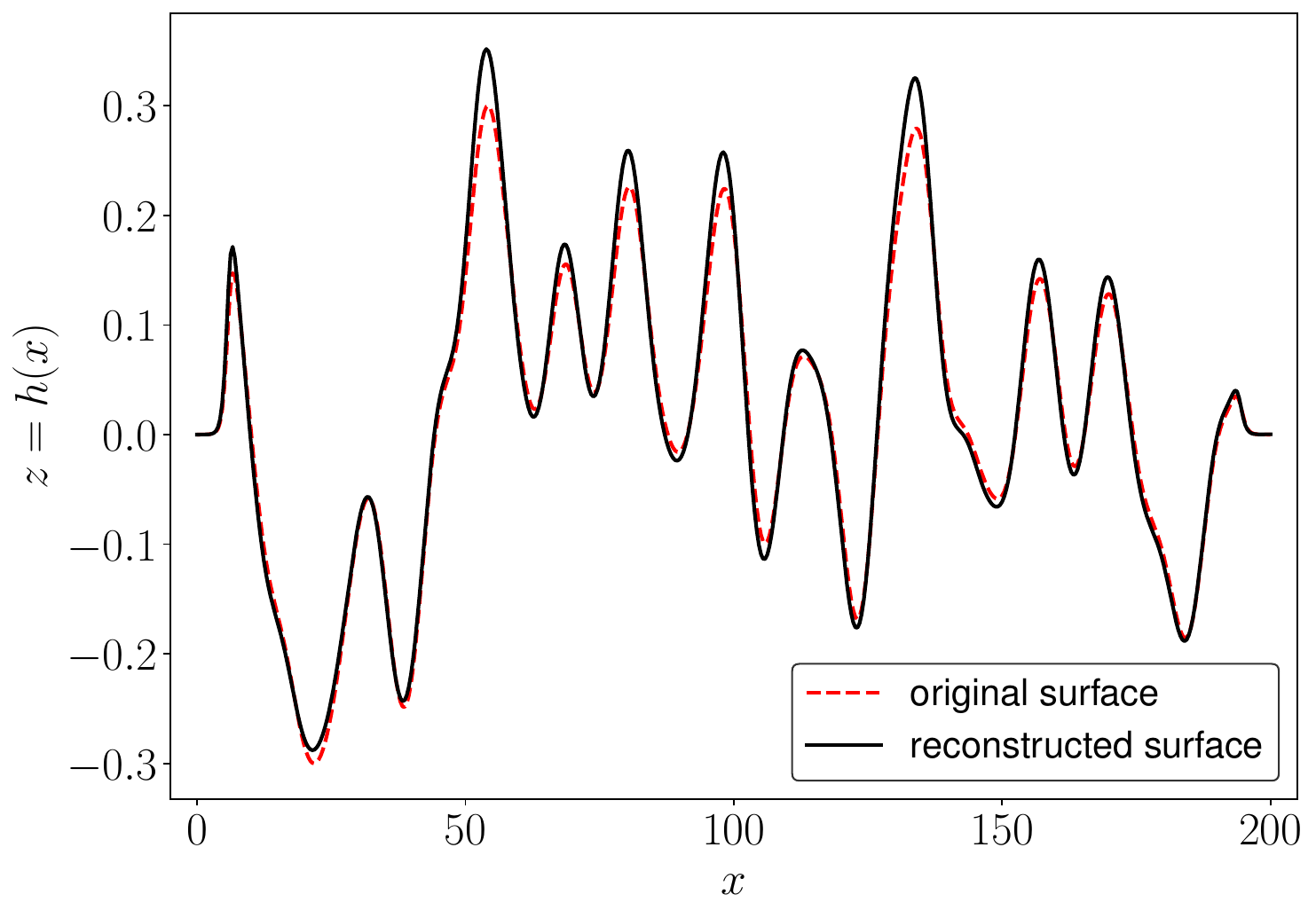}
  \caption{}
  \end{subfigure}
  \begin{subfigure}{0.46\textwidth}
  \centering
  \includegraphics[width=0.95\linewidth]{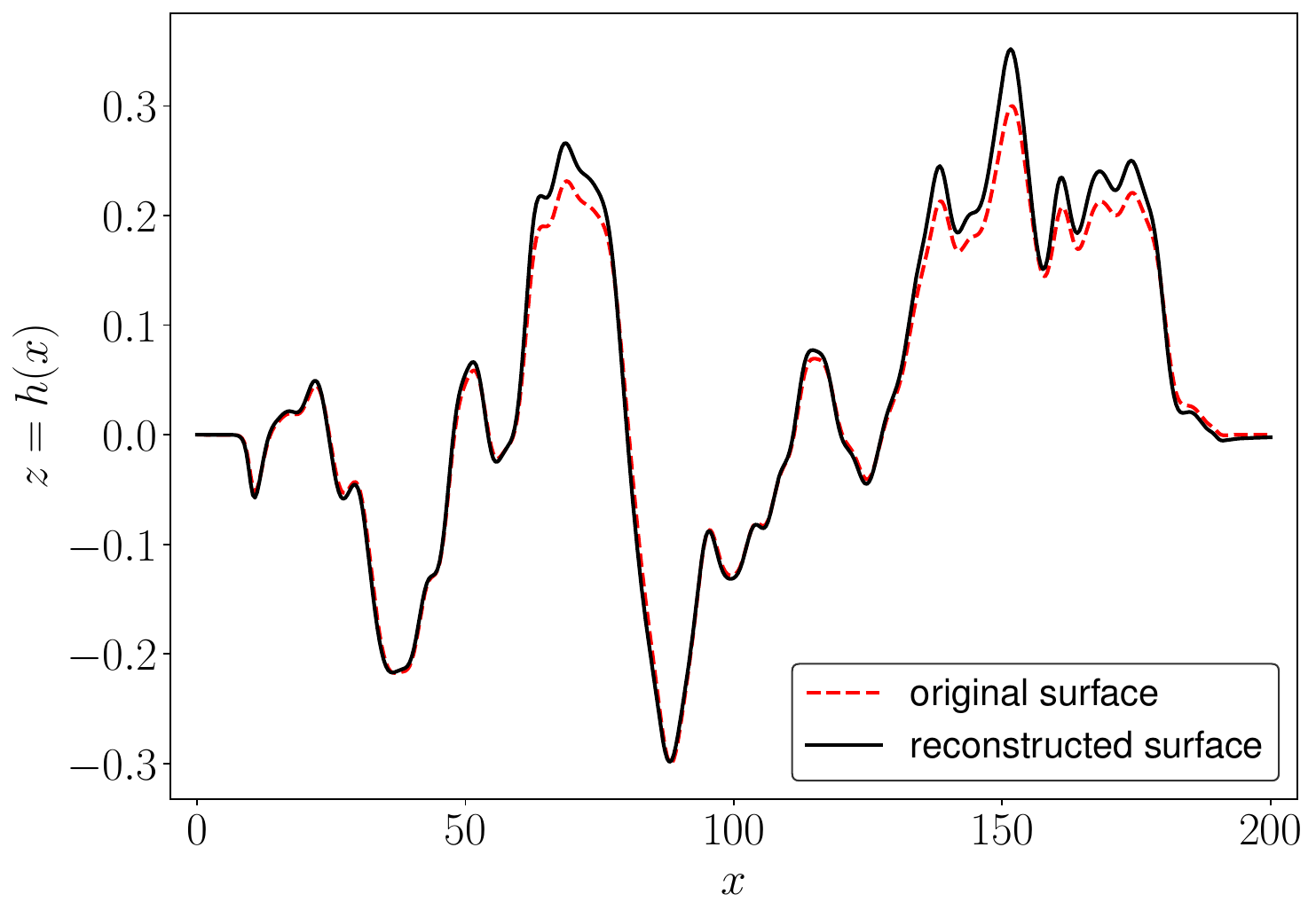}
  \caption{}
  \end{subfigure}
  \caption{Reconstruction of rough surfaces with Dirichlet boundary
  condition for surfaces with (a) Gaussian type and
  (b) sub--fractal autocorrelation functions.}
  \label{fig:diri_result}
\end{figure}

We first assume Dirichlet boundary conditions.
The discretization size is chosen to be $N =
500$. The receiver is placed at height $\zeta = 0.7$, which is
about $0.11\lambda$.
The horizontal surface scale of the roughness $l$ is initially set as the
same as the wavelength, although this will later be varied. Initially
the measurements are assumed to be noise-free. In the absence of noise
the reconstructed surface against the exact surface for both types of
surfaces is plotted in \cref{fig:diri_result}.  Clearly, the
reconstructed surface agrees closely with the exact surface. A small height
discrepancy can be seen between the reconstruction and exact surface
near the peaks, but all key features are well-captured and correctly
located horizontally.

\Cref{table:error_diri} gives the $\ell_2$-norm error between the
actual and recovered surfaces with different discretization sizes $N$
and different receiver heights $\zeta$ for the fixed surface scale
$l=\lambda$.
\begin{table}[htbp]
\begin{tabular}{c|ccc|ccc}
\hline
& \multicolumn{3}{c|}{Gaussian type surface} & \multicolumn{3}{c}{sub--fractal type surface} \\
\hline
& $\zeta = 0.7$ & $\zeta = 1.0$ & $\zeta = 1.2$ & $\zeta = 0.7$ & $\zeta= 1.0$ & $\zeta = 1.2$ \\
\hline
$N = 300$ ($\sim$9 nodes/$\lambda$) &0.143 & 0.204 & 0.281 & 0.158 & 0.214 & 0.303 \\
$N = 500$ ($\sim$15 nodes/$\lambda$) & 0.129 & 0.187 & 0.223 & 0.130 & 0.199 & 0.252 \\
$N= 800$ ($\sim$25 nodes/$\lambda$) & 0.112 & 0.151 & 0.222 & 0.112 &0.153 & 0.239 \\
\hline
\end{tabular}
\caption{The $\ell_2$ norm error between the actual and recovered
surfaces for the Dirichlet boundary condition with different number of
discretized points $N$ and different heights of the receiver $\zeta$.}
\label{table:error_diri}
\end{table}
Clearly, with more nodes used, the algorithm is more effective. The
receiver at height $\zeta =0.7$ gives the best result. The error is
slightly larger for the sub--fractal type surface.

The $\ell_2$-norm error for different values of surface scales ($l$)
is listed in \cref{table:error_l_diri} with fixed $N=500$ and $\zeta
= 0.7$.
\begin{table}[htbp]
\centering
\begin{tabular}{l|cc}
\hline
$l$ (scale size) & Gaussian type & sub--fractal \\
\hline
$l=0.5\lambda$ & 0.156 & 0.171 \\
$l=0.8\lambda$ & 0.130 & 0.144 \\
$l=\lambda$    & 0.129 & 0.130 \\
$l=1.2\lambda$ & 0.118 & 0.132 \\
$l=1.5\lambda$ & 0.114 & 0.136 \\
\hline
\end{tabular}
\caption{The $\ell_2$-norm error between the actual surface and the
recovered surface obtained using different surface scale sizes $l$
and fixing $N=500$ and $\zeta = 0.7$ under the Dirichlet boundary condition.}
\label{table:error_l_diri}
\end{table}
As the surface scale decreases, more peaks occur within the
reconstruction domain, which leads to a mild increase in error.

\begin{figure}
\centering
  \begin{subfigure}{0.6\textwidth}
  \centering
  \includegraphics[width=0.95\linewidth]{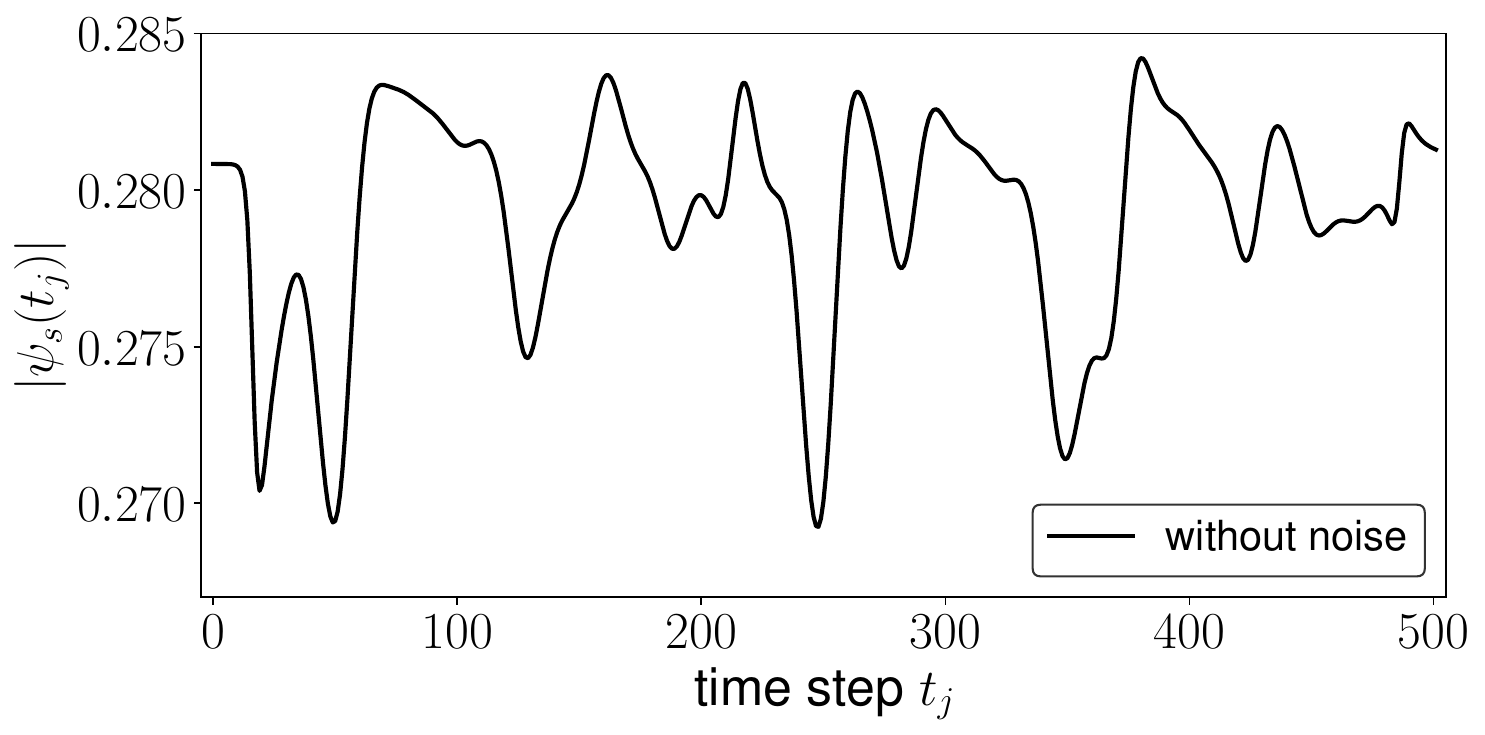}
  \end{subfigure}
  \begin{subfigure}{1.0\textwidth}
  \centering
  \includegraphics[width=0.6\linewidth]{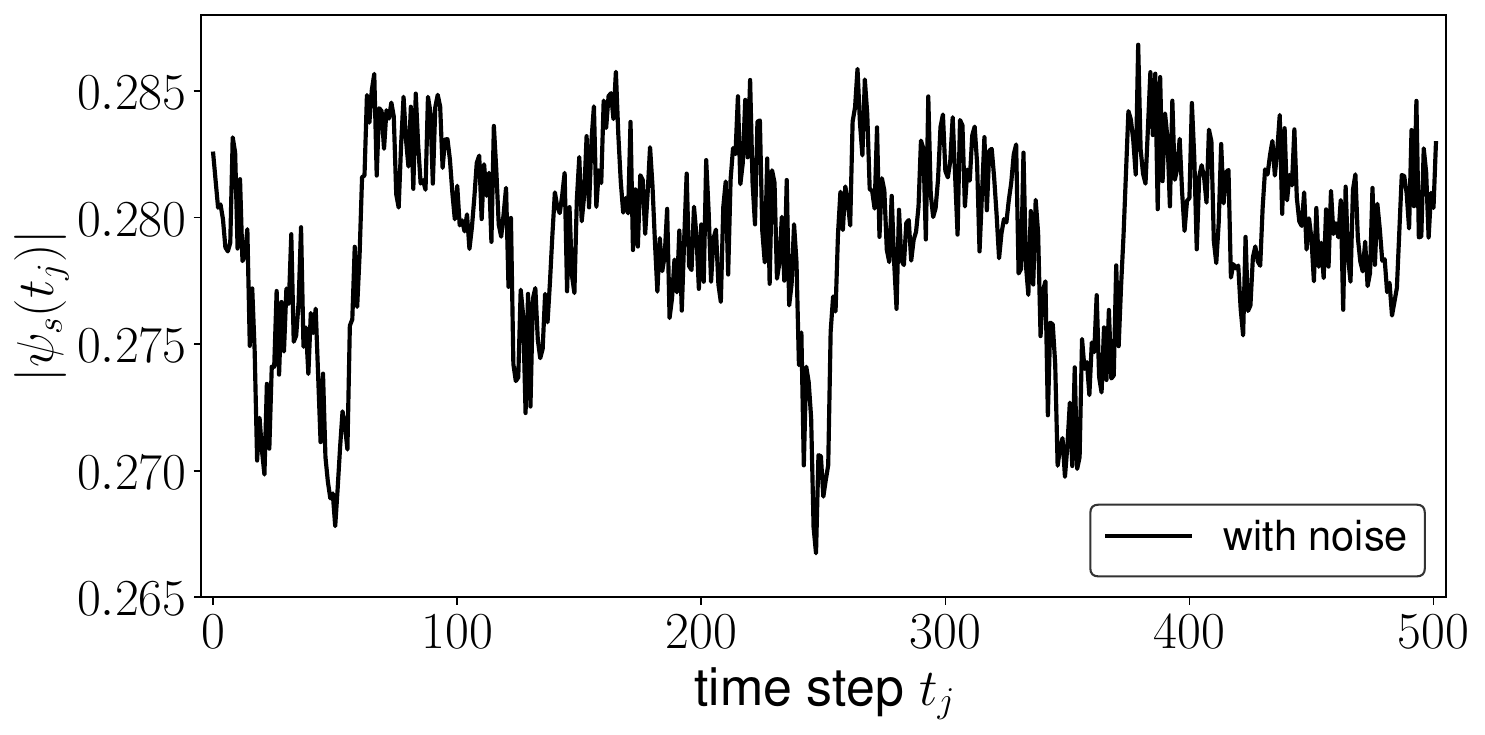}
  \end{subfigure}
  \caption{Amplitude of scattered data at each time step without
  noise (upper) and with $3\%$ noise (lower) under Dirichlet boundary condition.}
  \label{fig:diri_data_noise}
\end{figure}

\begin{figure}
\centering
  \begin{subfigure}{0.46\textwidth}
  \centering
  \includegraphics[width=0.95\linewidth]{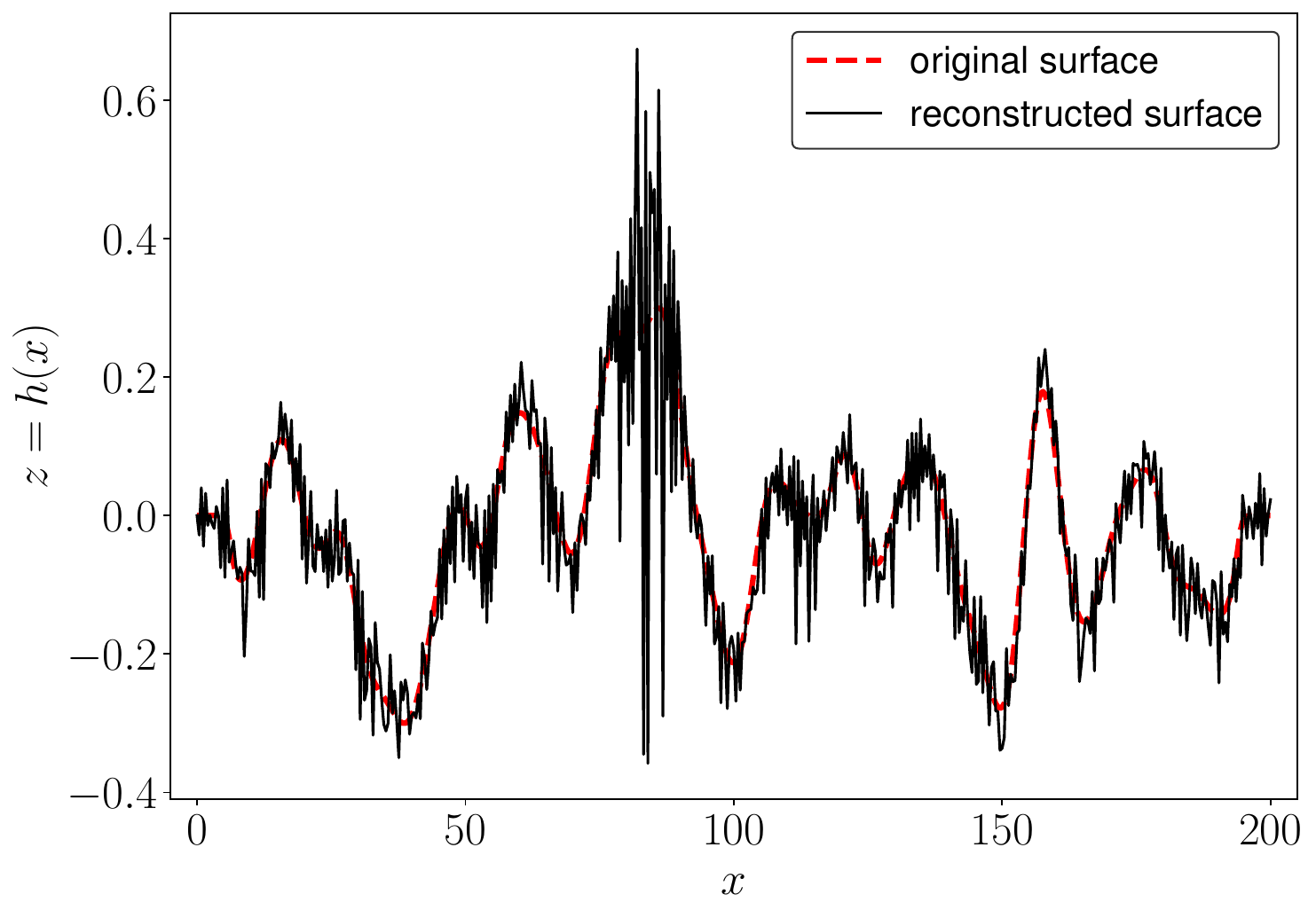}
  \caption{}
  \end{subfigure}
  \begin{subfigure}{0.46\textwidth}
  \centering
  \includegraphics[width=0.95\linewidth]{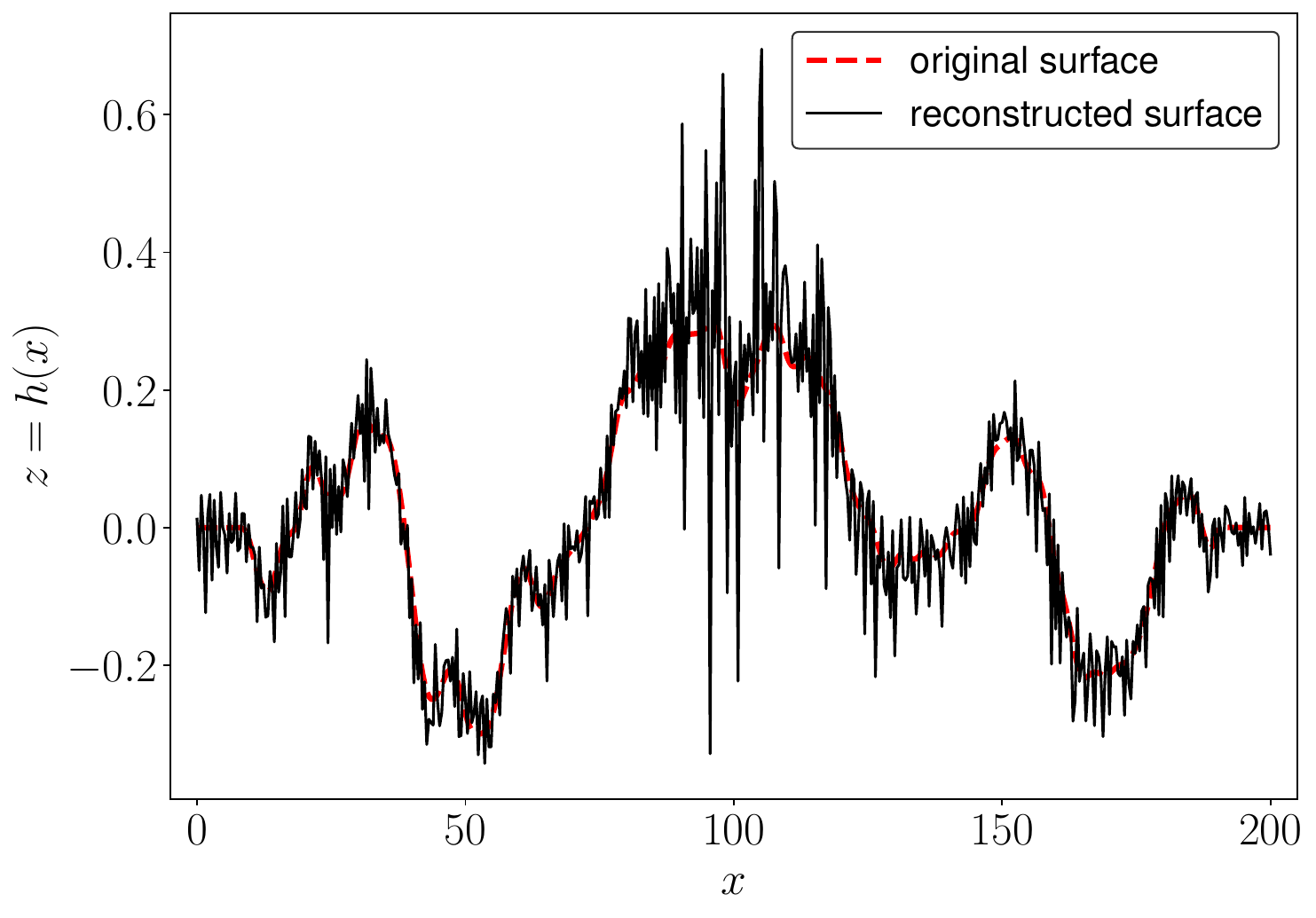}
  \caption{}
  \end{subfigure}
  \caption{Reconstruction of rough surfaces with Dirichlet boundary
  condition using $3\%$ noisy scattered data for
  (a) Gaussian type surface and (b) sub--fractal type surface.}
  \label{fig:diri_noise}
\end{figure}

\begin{figure}
\centering
  \begin{subfigure}{0.46\textwidth}
  \centering
  \includegraphics[width=0.95\linewidth]{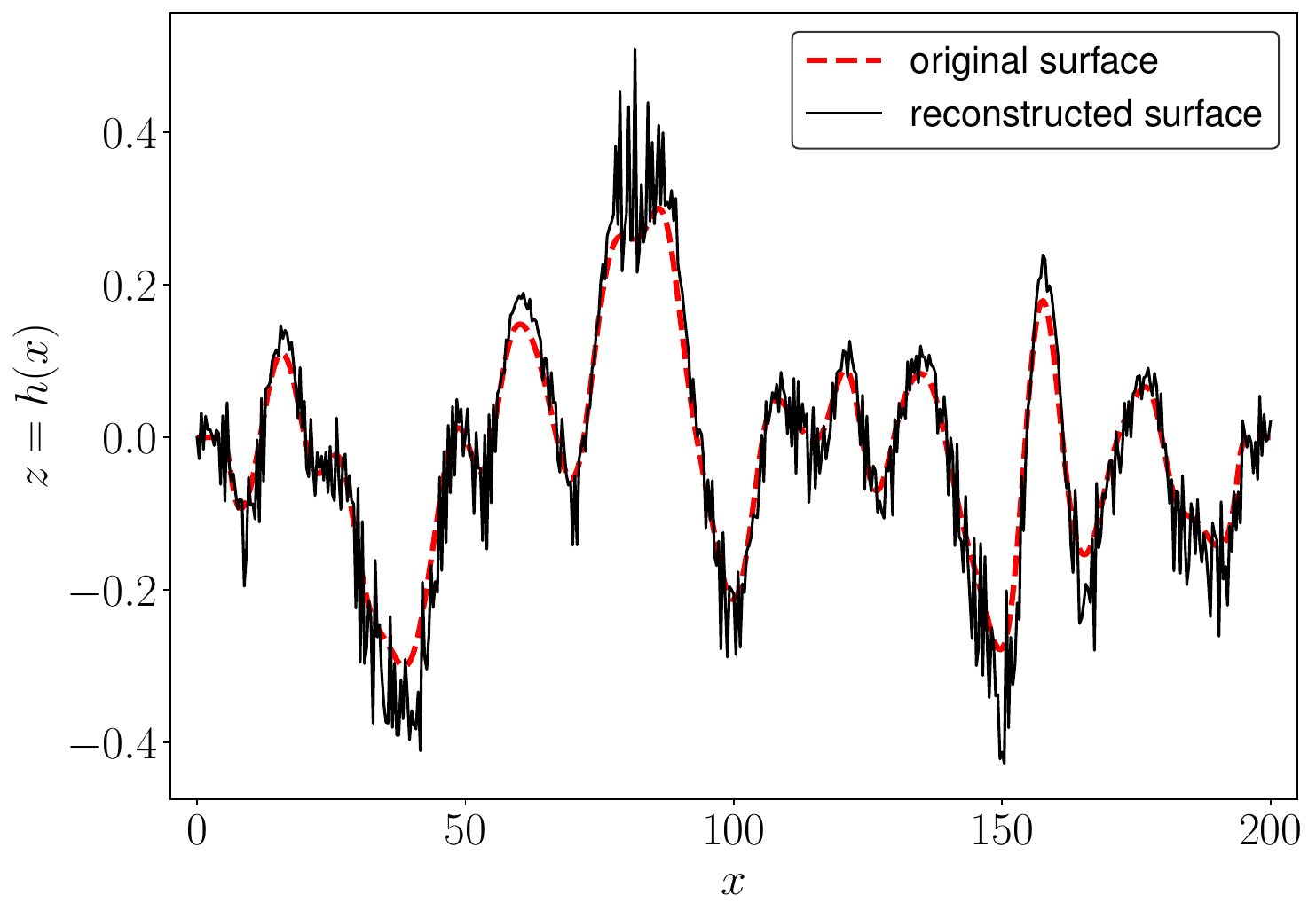}
  \caption{}
  \end{subfigure}
  \begin{subfigure}{0.46\textwidth}
  \centering
  \includegraphics[width=0.95\linewidth]{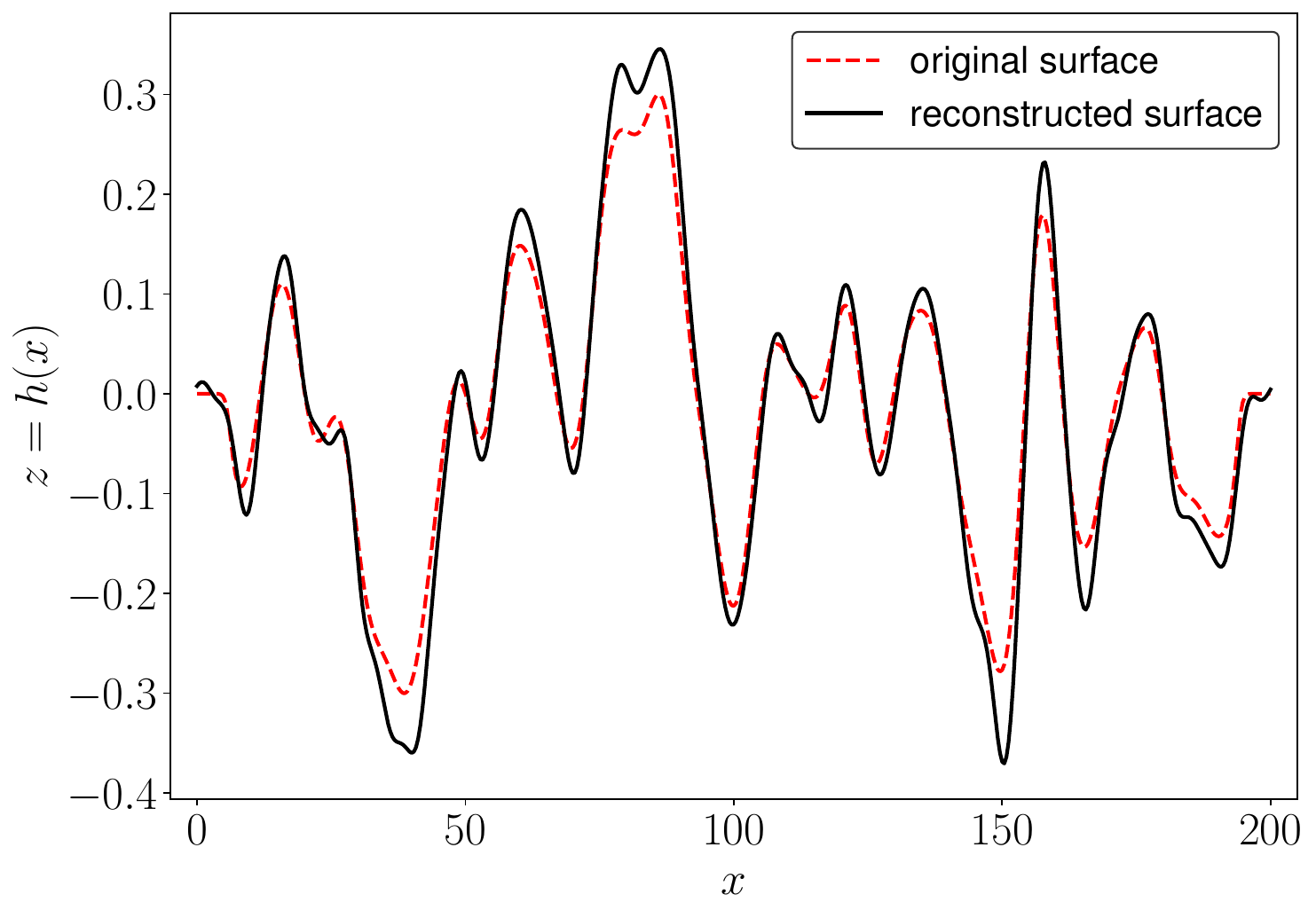}
  \caption{}
  \end{subfigure}
  \hfill
  \begin{subfigure}{0.46\textwidth}
  \centering
  \includegraphics[width=0.95\linewidth]{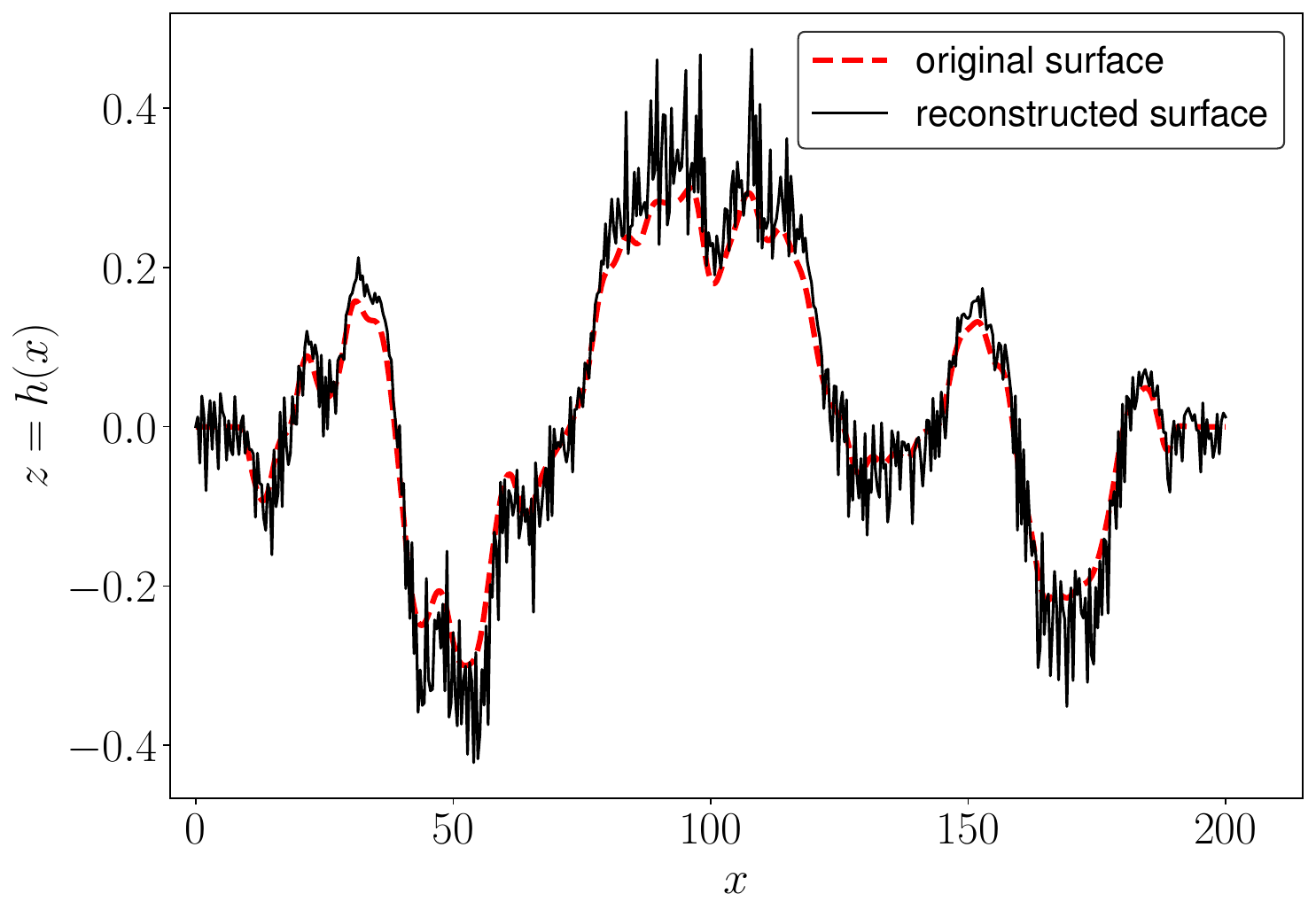}
  \caption{}
  \end{subfigure}
  \begin{subfigure}{0.46\textwidth}
  \centering
  \includegraphics[width=0.95\linewidth]{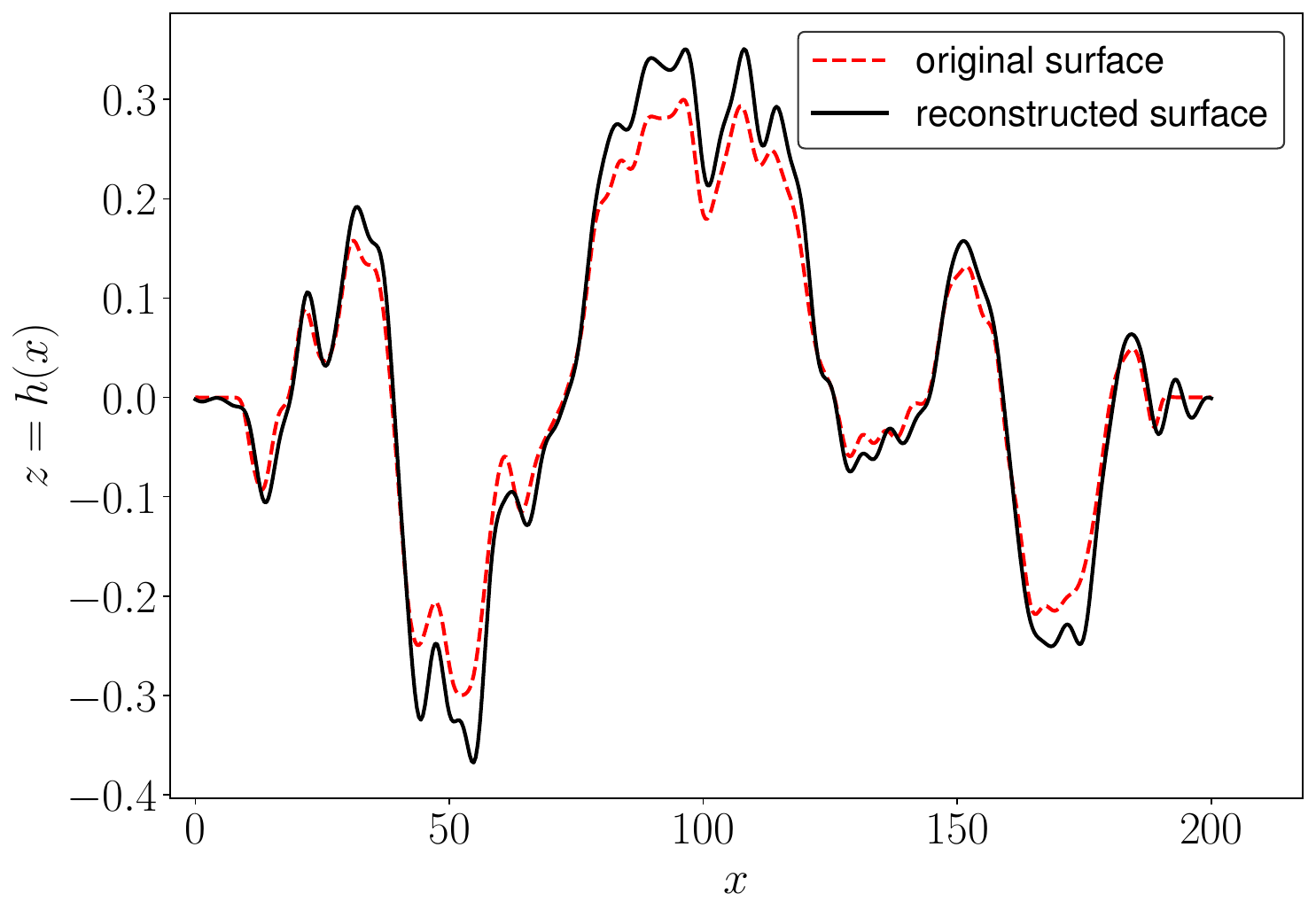}
  \caption{}
  \end{subfigure}
  \caption{Reconstruction of rough surfaces under Dirichlet boundary
  condition with $3\%$ noisy scattered data using the derivative filter
  in the algorithm for (a) Gaussian type surface and (b) sub--fractal
  type surface; plus an additional FFT--type filter
  for (c) Gaussian type surface and (d) sub--fractal
  type surface.}
  \label{fig:diri_filter}
\end{figure}

In order to test the robustness of the approach, we now add $3\%$
noise to the scattered data at each time step, and the resulting 
amplitude of scattered field with and without noise is shown in
\cref{fig:diri_data_noise} 
using $N=500$, $\zeta = 0.7$, and $l=\lambda$ for a Gaussian type surface.
With the noisy data, the recovered surfaces compared to the actual
surfaces are shown in \cref{fig:diri_noise}.
With the noisy data, the reconstructed surfaces capture all main
features, but before further processing display fine scale
fluctuations throughout the 
domain, which for the most part are qualitatively similar to the measurement
noise. However a few much larger spikes appear.
Their presence is due to the ambiguity of taking the logarithm in
\cref{eqn:diri_h1}. On the other hand, 
it is observed that there are only a small number of these spikes in
each case.  
To eliminate this problem, we introduce the following
post-reconstruction `filter' into the algorithm: Suppose that the surface 
heights recovered at the time steps $t_{j-1}$ and $t_{j}$ are
$H_{j-1}$ and $H_j$ obtained 
by \cref{eqn:diri_h1,eqn:diri_h2}. We force the surface height $H_j$
equal to $H_{j-1}$ 
if the surface derivative at this point is larger than one, namely
$\left \lvert (H_j-H_{j-1})/\Delta x \right \rvert > 1$. With this
filter, 
the reconstructed surfaces are shown in \cref{fig:diri_filter}.
A straightforward FFT--type filter can be applied to damp out most oscillations.
After filtering, the reconstructed surface recaptures the overall
shape of the original surface. We also list the $\ell_2$--norm error
between the filtered recovered surfaces and the actual surfaces in
\cref{table:error_diri_noise}. 
\begin{table}[htbp]
\centering
\begin{tabular}{c|cc}
\hline
\multicolumn{3}{c}{$\ell_2$-norm error}        \\
\hline
noise level  & Gaussian type surface & sub--fractal type surface \\
\hline
2\%           & 0.189                 & 2.231                    \\
3\%           & 0.219                 & 0.258                    \\
5\%            & 0.245                 & 0.276                    \\
7\%            & 0.279                 & 0.297                    \\
8\%            & 0.311                 & 0.342                   \\
\hline
\end{tabular}
\caption{The $\ell_2$-norm error between the actual surface and the
filtered recovered surface obtained by suing $N=500$, $\zeta = 0.7$
and $l=\lambda$ 
under the Dirichlet boundary condition with noisy scattered data of
different noise levels.} 
\label{table:error_diri_noise}
\end{table}
The $\ell_2$--norm error increases as the noise level goes up. The
error is larger 
for the recovered sub--fractal surface.
%-----------------------------------------------------------------------------
\subsection{Neumann boundary condition}
\begin{figure}
  \centering
  \begin{subfigure}{0.46\textwidth}
    \centering
    \includegraphics[width=0.95\linewidth]{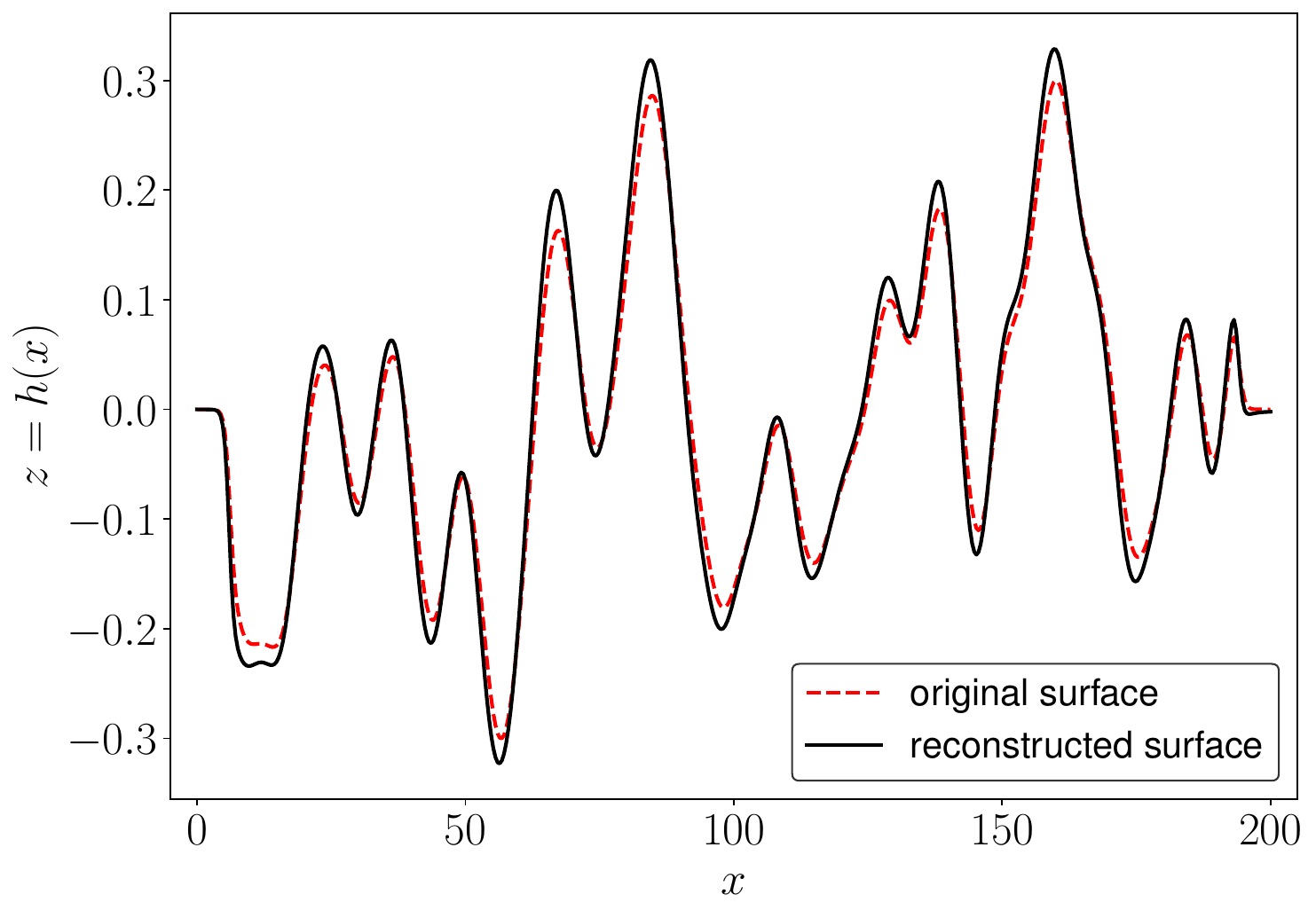}
    \caption{}
  \end{subfigure}
  \begin{subfigure}{0.46\textwidth}
    \centering
    \includegraphics[width=0.95\linewidth]{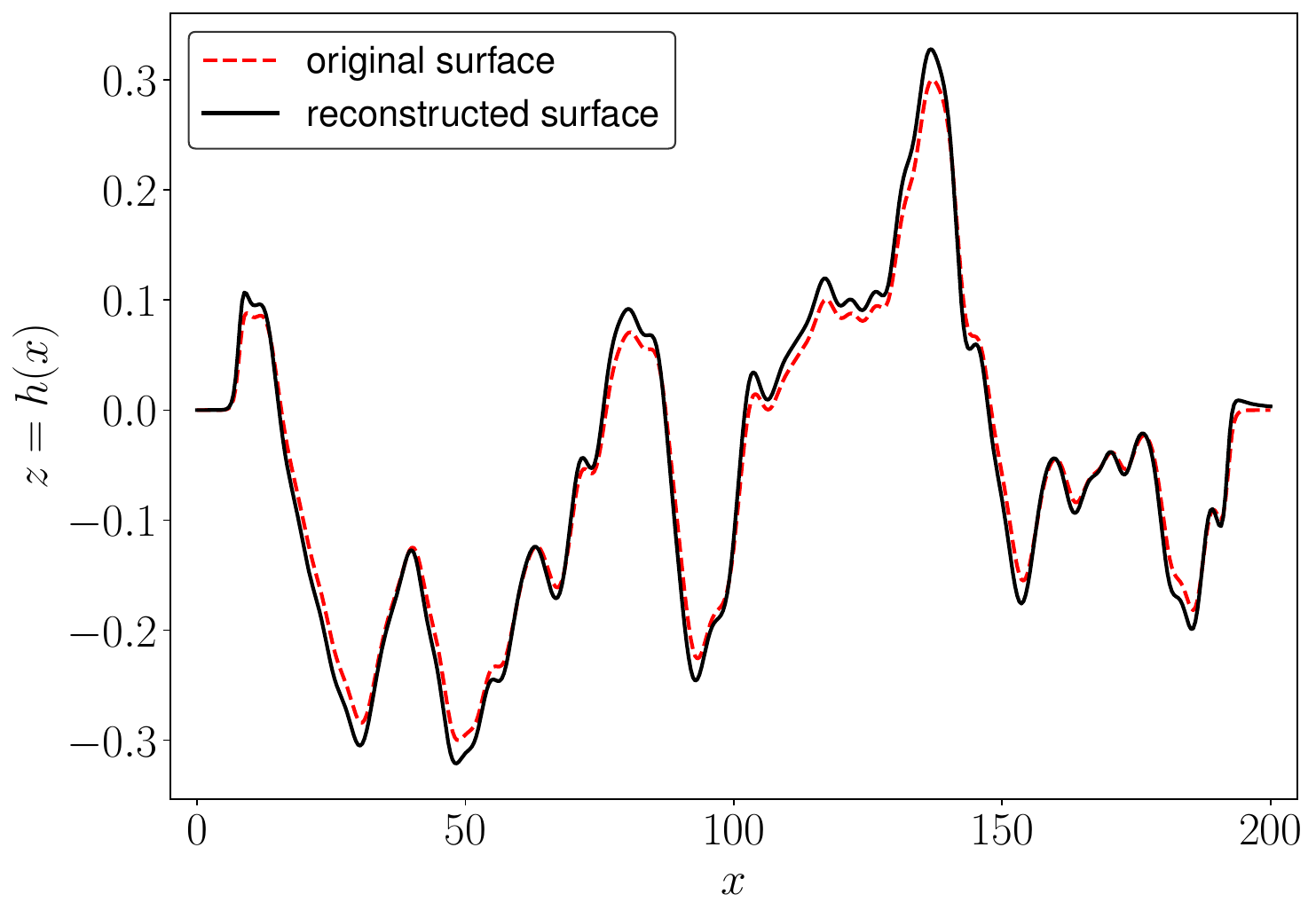}
    \caption{}
  \end{subfigure}
  \caption{Reconstruction of rough surfaces with Neumann boundary condition
  for surfaces of (a) Gaussian type and (b) sub--fractal type.}
  \label{fig:neu_result}
\end{figure}

The applicability of the approach for Neumann boundary condition is
now examined. 
With the  discretization size $N=500$, surface scale $l = \lambda$, and the
receiver height $\zeta = 0.7$, the reconstructed surface versus the
true surface is plotted in \cref{fig:neu_result} for Gaussian and
sub--fractal type surfaces under the Neumann boundary condition, in
the absence of measurement noise.
Similar to the Dirichlet case, the recovered surface closely matches the
original surface, with the largest errors apparent around the peaks.

Fix the surface scale $l=\lambda$, the $\ell_2$-norm error for using
different sizes of discretization and heights of receiver for both
types of surfaces is summarised in \cref{table:error_neu}.
\begin{table}[htbp]
\begin{tabular}{c|ccc|ccc}
\hline
& \multicolumn{3}{c|}{Gaussian type surface} & \multicolumn{3}{c}{sub--fractal type surface} \\
\hline
& $\zeta =0.7$ & $\zeta = 1.0$ & $\zeta = 1.2$ & $\zeta = 0.7$ & $\zeta = 1.0$ & $\zeta = 1.2$ \\
\hline
$N = 300$ ($\sim$9 nodes/$\lambda$) & 0.148 & 0.159 & 0.311 & 0.148 & 0.151 & 0.322 \\
$N = 500$ ($\sim$15 nodes/$\lambda$) & 0.135 & 0.136 & 0.257 & 0.133 & 0.133 & 0.252 \\
$N = 800$ ($\sim$25 nodes/$\lambda$) & 0.120 & 0.131 & 0.241 & 0.125 & 0.128 & 0.237 \\
\hline
\end{tabular}
\caption{The $\ell_2$ norm error between the actual and recovered
surfaces under the Neumann boundary condition with different number of
discretized points $N$ and different height of the receiver $\zeta$.}
\label{table:error_neu}
\end{table}
The size of error is similar to the Dirichlet case.
It is found that the method is robust with respect to the
discretization size. Within this range, and with a larger number of points
used, the error reduces.
In the Neumann case, the error for both recovered Gaussian type
and sub--fractal surfaces keeps at the similar level.
The heights of the receiver at $\zeta = 0.7$ and $\zeta=1.0$ give
similar results while the error increases for $\zeta = 1.2$.

Now, we set $N=500$ and $\zeta = 0.7$, the error with respect to different
values of surface scales ($l$) is listed in \cref{table:error_l_neu}.
\begin{table}[htbp]
\centering
\begin{tabular}{l|cc}
\hline
$l$ (scale size)  & Gaussian type & sub-fractal \\
\hline
$l=0.5\lambda$   & 0.149         & 0.149       \\
$l=0.8\lambda$   & 0.121         & 0.137       \\
$l=\lambda$      & 0.135         & 0.130       \\
$l=1.2\lambda$   & 0.128         & 0.125       \\
$l=1.5\lambda$   & 0.122         & 0.122       \\
\hline
\end{tabular}
\caption{The $\ell_2$-norm error between the actual and the
recovered surfaces obtained by using different surface scale sizes $l$
and fixing $N=500$ and $\zeta = 0.7$ under the Neumann boundary condition.}
\label{table:error_l_neu}
\end{table}
Unsurprisingly, the performance degrades as the surface becomes more
oscillatory for smaller $l$.

\begin{figure}
  \centering
  \begin{subfigure}{0.6\textwidth}
  \centering
  \includegraphics[width=0.95\linewidth]{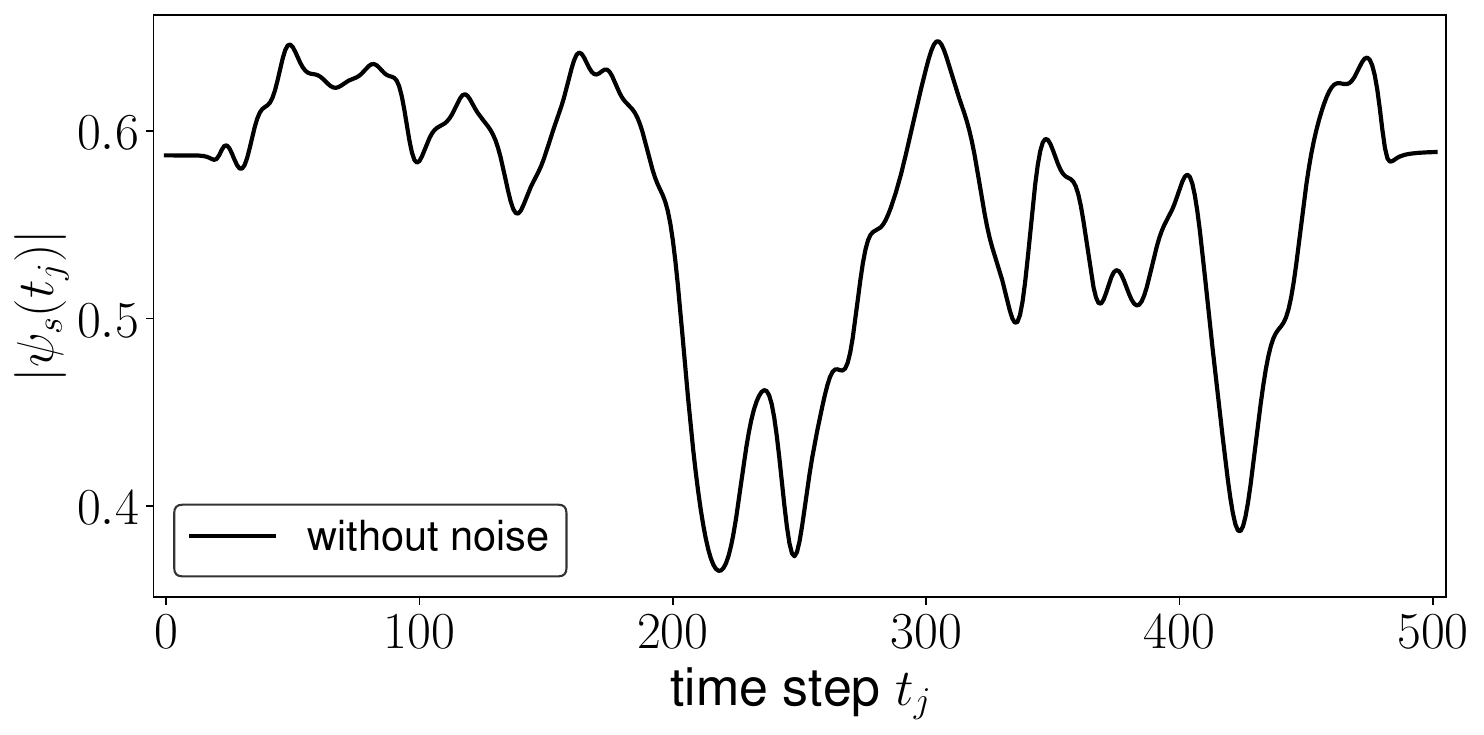}
  \end{subfigure}
  \begin{subfigure}{1.0\textwidth}
  \centering
  \includegraphics[width=0.6\linewidth]{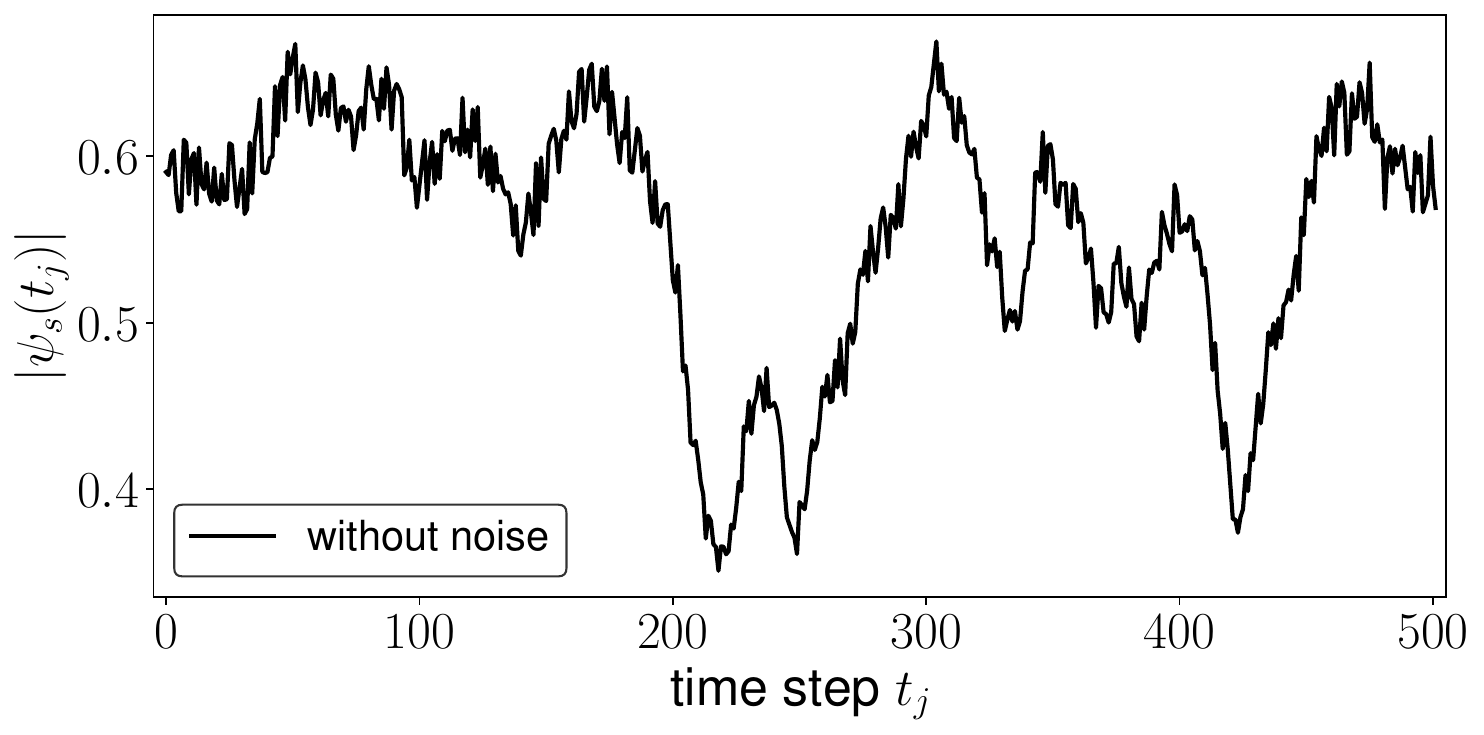}
  \end{subfigure}
  \caption{Amplitude of scattered data at each time step without noise (upper)
  and with $3\%$ noise (lower) under Neumann boundary condition.}
  \label{fig:neu_data_noise}
\end{figure}

\begin{figure}
\centering
  \begin{subfigure}{0.46\textwidth}
  \centering
  \includegraphics[width=0.95\linewidth]{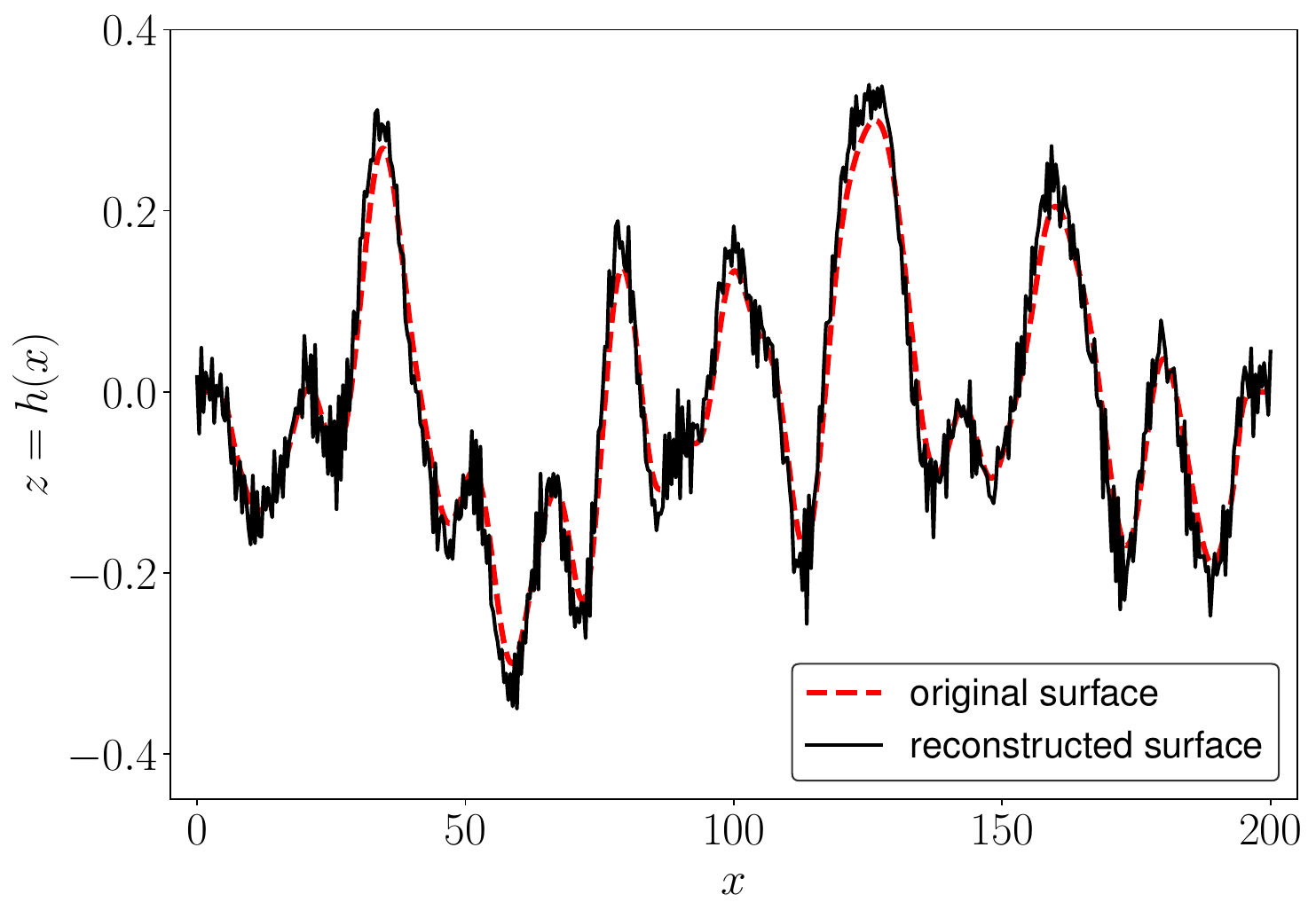}
  \caption{}
  \end{subfigure}
  \begin{subfigure}{0.46\textwidth}
  \centering
  \includegraphics[width=0.95\linewidth]{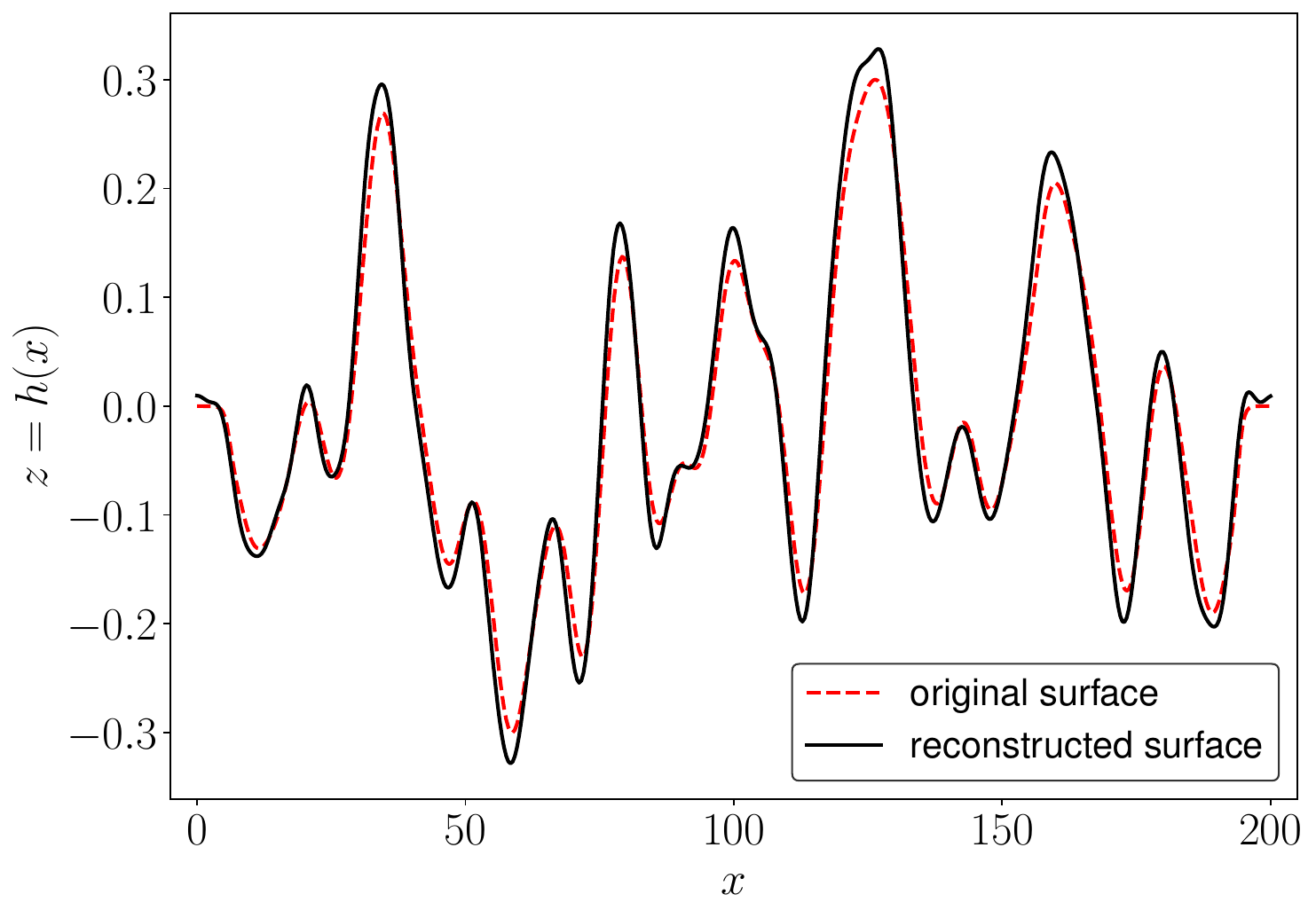}
  \caption{}
  \end{subfigure}
  \hfill
  \begin{subfigure}{0.46\textwidth}
  \centering
  \includegraphics[width=0.95\linewidth]{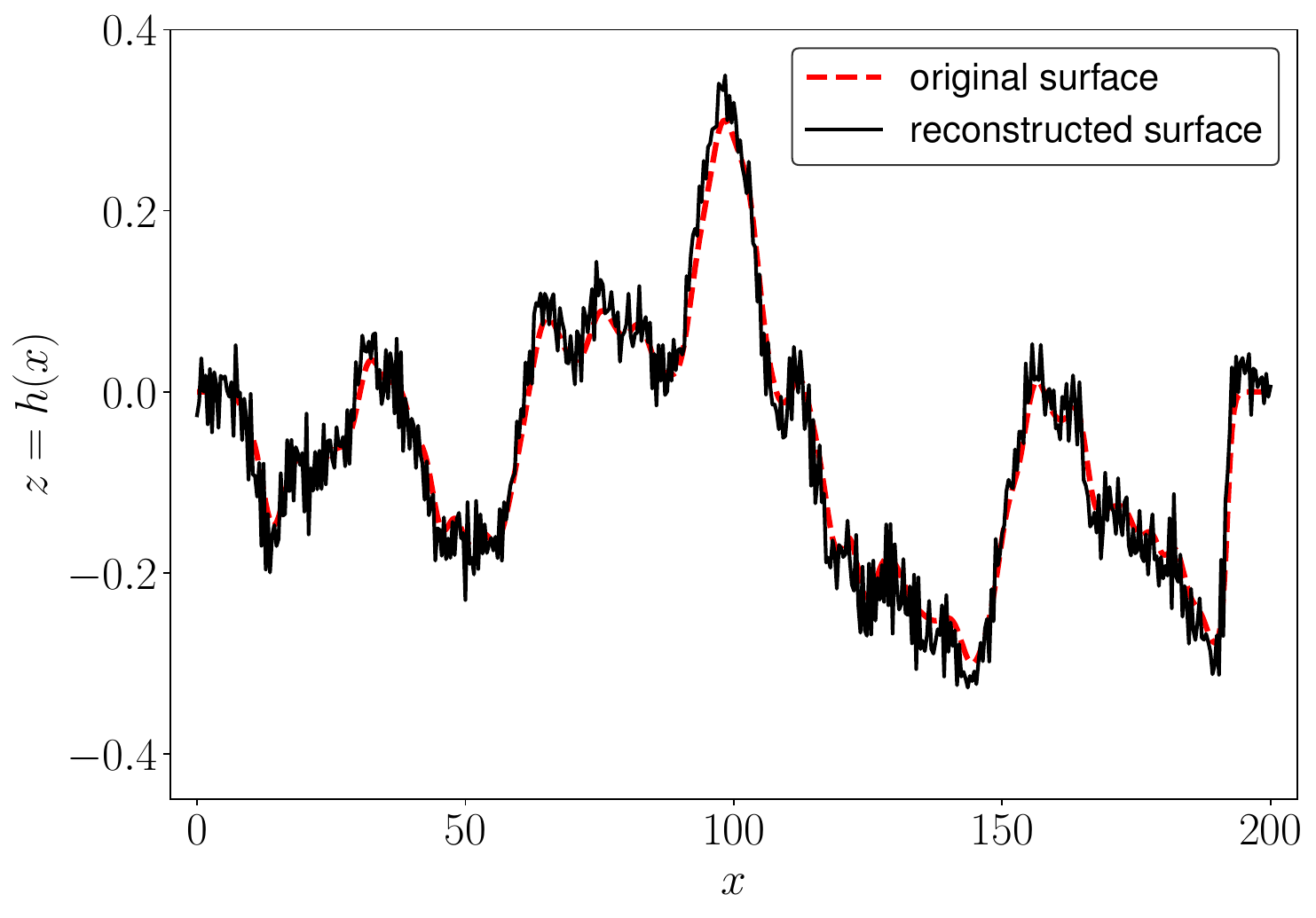}
  \caption{}
  \end{subfigure}
  \begin{subfigure}{0.46\textwidth}
  \centering
  \includegraphics[width=0.95\linewidth]{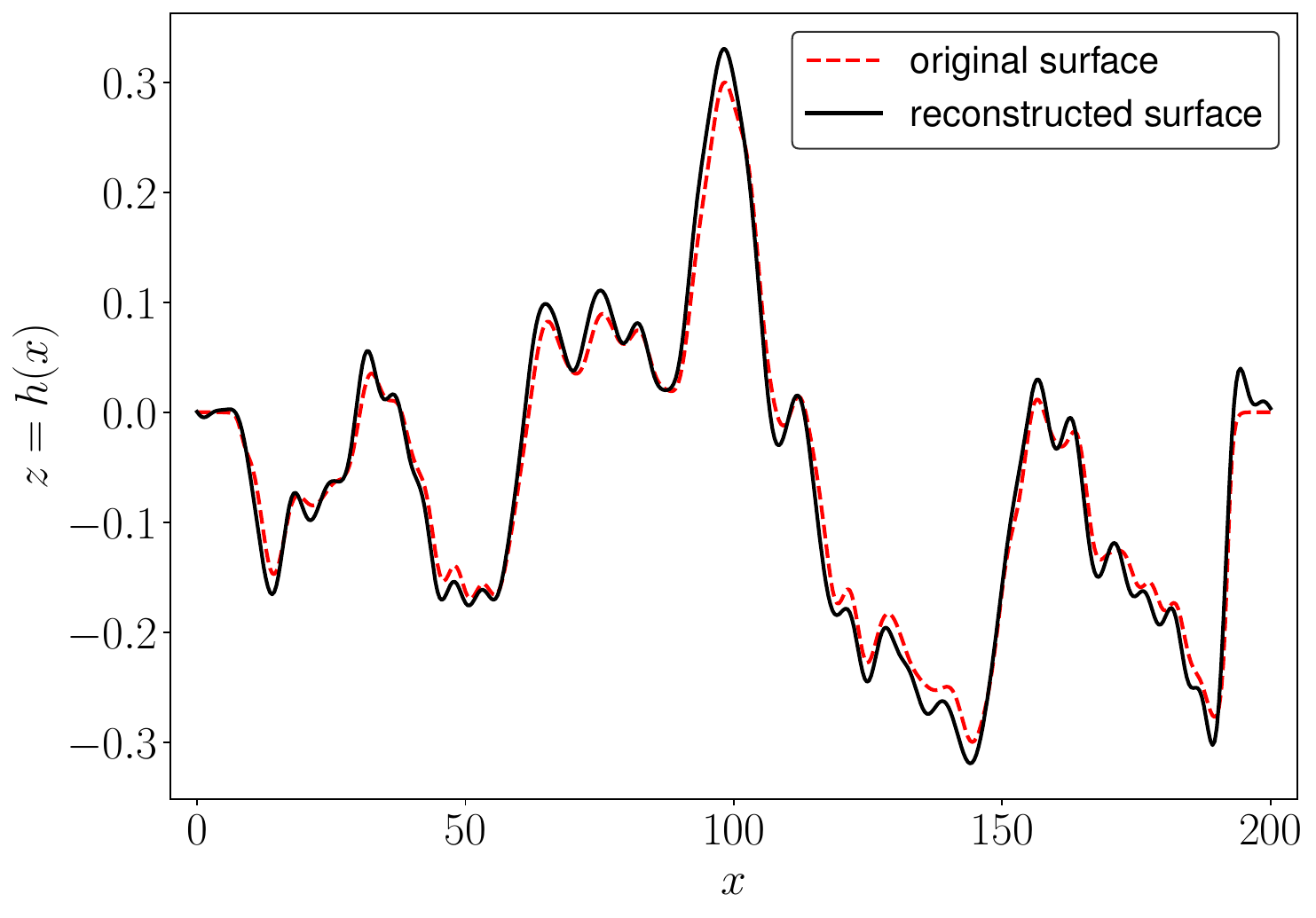}
  \caption{}
  \end{subfigure}
  \caption{Reconstruction of rough surfaces under Neumann boundary
  condition with $3\%$ noisy scattered data for (a) Gaussian type
  surface and (b) sub--fractal type surface; and filtered recovered
  surfaces of (c) Gaussian type and (d) sub--fractal type
  using a FFT type filter.}
  \label{fig:neu_noise}
\end{figure}

A white noise is added to the scattered data at each time step with the noise level
$3\%$, the magnitude of the data with and without noise is shown in \cref{fig:neu_data_noise}
with $N=500$, $\zeta = 0.7$ and $l=\lambda$ for sub--fractal type surface.
The reconstruction results for both type surfaces with noisy data are
shown in \cref{fig:neu_noise}. 
The method for the Neumann boundary condition is more robust with respect
to the noise compared to the Dirichlet case. Without a filter, the
reconstructed surface, 
though contains certain oscillations, still closely matches the actual surface.
For the sub-fractal case, while the solution clearly captures the main
surface peaks and troughs, the noise initially masks the
smaller scale features.
However, with a straightforward FFT type filter,
the reconstruction can capture most shapes of the true surface.
Finally, the $\ell_2$-norm error between the filtered rough surface
and the actual surface 
of two different types with the noise data is listed in
\cref{table:error_neu_noise}.
\begin{table}[htbp]
\centering
\begin{tabular}{c|cc}
\hline
\multicolumn{3}{c}{$\ell_2$-norm error}        \\
\hline
noise level  & Gaussian type surface & sub--fractal type surface \\
\hline
2\%            & 0.137                 & 0.138                   \\
3\%            & 0.137                 & 0.140                    \\
5\%            & 0.141                 & 0.144                    \\
7\%            & 0.177                 & 0.185                   \\
8\%            & 0.220                 & 0.256                   \\
\hline
\end{tabular}
\caption{The $\ell_2$-norm error between the actual surface and the
filtered recovered surface obtained by suing $N=500$,$\zeta = 0.7$ and
$l=\lambda$ under the Neumann boundary condition with noisy
data of different noise levels.}
\label{table:error_neu_noise}
\end{table}
If the noise level is not big enough ($\geq 5\%$), the error of
recovered surface using noisy data keeps at the same level with
non-noisy data under the Neumann boundary condition. As noise level
increases more, the increment in the error is clear.

%-----------------------------------------------------------------------------
\section{Conclusions}
\label{sec:conclusions}

A novel reconstruction algorithm has been developed to recover the
rough surface based on repeated measurement of wave field at a single
receiver. The surface is moved with respect to source and
receiver during the algorithm, and the method
is of a marching--type, in which the surface height is recovered
progressively as the position changes. This approach is based on
the parabolic wave integral equations. Both Dirichlet and Neumann
boundary conditions have been tackled.  The solution relies on low
grazing angle. Although the parabolic equation Green's function which
neglects backscattering, a similar approach using the full Green's
function is feasible. 

Through numerical examples, it was found that the solution surface
captures most shapes of the original surface for both Gaussian and
sub--fractal type surfaces. Good agreement between recovered and
real surface was observed, whereas large error shows up around peaks
of surface. The results with respect to random white noise was also
tested. Though some instabilities are present in the recovered surface
under the Dirichlet boundary condition, the large peak--type errors
can be damped out completely by employing a derivative filter in the
algorithm. The noise induced in the solution is qualitatively similar
to the measurement noise, i.e. it is delta-correlated. With a FFT-type
filter, the reconstructed surface closely matches the exact solution.

A main motivation for this work is that the
principles of the method can be extended in 
two important directions. Work is
currently underway to adapt the approach to phaseless
data. If the measurement receiver gives the amplitude of the
wave field
the reconstruction algorithm may be far more challenging. Some related
previous work can be found in \cite{rough2}, but in which the surface
is stationary and large number of receivers are required. In addition,
the principles of this method are feasibly applicable to
three--dimensional (3D) problems. In 3D, applying the standard marching method
over a square $N\times N$ domain (say) would require $O(N^2)$
receivers to obtain 
sufficient scattered data. Such a receiver array would in general be both
difficult to accomplish and impractical without self-interference.
The approach here would replace this with a source and line array,
moving with respect to the surface. At each step of the marching
iteration the surface would be recovered along a line, orthogonal to
the direction of the Gaussian beam.  

%-----------------------------------------------------------------------------
\section*{Acknowledgements}

The first author acknowledges supports from the the Youth Program of the Natural
Science Foundation of Jiangsu Province (No.~BK20230466),
the Jiangsu Funding Program for Excellent Postdoctoral Talent (No.~2022ZB584),
and Jiangsu Shuangchuang Project (JSSCTD202209).
%-----------------------------------------------------------------------------
%\bibliographystyle{unsrtnat}
\bibliographystyle{unsrt}
\bibliography{references}
%-----------------------------------------------------------------------------
\end{document}